\documentclass[a4paper,11pt]{article}
\pdfoutput=1 

\usepackage{jheppub} 

\usepackage[T1]{fontenc}
\usepackage{amsfonts}
\usepackage{amssymb}
\usepackage[utf8]{inputenc} 
\usepackage{graphicx} 
\usepackage{braket}
 \usepackage{fancyhdr}
\usepackage{mathrsfs}
\usepackage{setspace}
\usepackage{amsfonts}
\usepackage{amssymb}
\usepackage{amsmath}
\usepackage{bbm}
\usepackage{epsfig}
\usepackage{latexsym}
\usepackage{color}
\usepackage{nicefrac}
 \usepackage{slashed}
 \usepackage{multirow}
 \usepackage{comment}
 \usepackage{hyperref}
\usepackage{slashed}
\usepackage{array}
\usepackage{booktabs}
\usepackage{float}

\def\be{\begin{equation}}
\def\ee{\end{equation}}
\def\del{\partial}
\newcommand{\Tr}{\text{Tr}}

 \newcommand   \cO {\mathcal{O}}
\def\sn{{\rm sn}}
\def\cn{{\rm cn}}
\def\dn{{\rm dn}}
\def \ELE {{\mathbb{E}}}
\def \KK {{\mathbb{K}}}
\newcommand{\ba}{\begin{eqnarray}}
\newcommand{\ea}{\end{eqnarray}}

\def \T {{\cal T}}

\newcommand{\order}[1]{ \mathcal{O}\left({#1}\right) }
\newcommand{\abs}[1]{\left \lvert {#1} \right\rvert}
\title{\boldmath
Semiclassical Canovaccio for  Composite Operators 
}

\author{
Oleg Antipin$^{1}$,
Jahmall Bersini$^{2}$,
Jacob Hafjall$^{3}$,
Giulia Muco$^{3}$,
Francesco Sannino$^{3,4}$
}

\emailAdd{oantipin@irb.hr}
\emailAdd{jahmall.bersini@unibe.ch}
\emailAdd{jahaf21@student.sdu.dk}
\emailAdd{giulia@qtc.sdu.dk}
\emailAdd{sannino@qtc.sdu.dk}

\affiliation{
$^{1}$Rudjer Boskovic Institute, Division of Theoretical Physics,
Bijeni\v cka 54, 10000 Zagreb, Croatia
}

\affiliation{
$^{2}$Albert Einstein Center for Fundamental Physics,
Institute for Theoretical Physics, University of Bern,
Sidlerstrasse 5, CH-3012 Bern, Switzerland
}

\affiliation{
$^{3}$Quantum Theory Center ($\hslash$QTC) at IMADA \& D-IAS,
Southern Denmark University, Campusvej 55, 5230 Odense M, Denmark
}

\affiliation{
$^{4}$Dept. of Physics E. Pancini, Universit\`a di Napoli Federico II,
via Cintia, 80126 Napoli, Italy
}

\date{}
\abstract{  
We present a novel semiclassical framework tailored to determine the scaling dimensions of heavy neutral composite operators in conformal field theories (CFTs) which are inaccessible with other current methodologies. It utilizes the state-operator correspondence to map the desired scaling dimensions to the semiclassical energy spectrum of periodic homogeneous field configurations on a cylinder.  As concrete applications, we provide detailed analyses for the \(\phi^4\) theory near four dimensions and $\phi^6$ near three dimensions, semiclassically determining the full spectrum of neutral operators in the traceless symmetric Lorentz representations. Our methodology is presented pedagogically and is readily applicable to a vast class of CFTs. 
}

\arxivnumber{}

\begin{document}

\maketitle

\section{Introduction}

Conformal field theories (CFT)s describe a plethora of phase transitions, as established in pioneering works on criticality and the renormalization group \cite{Landau:1937obd, Kadanoff:1966wm, Kadanoff:1969zz, Wilson:1973jj, Wilson:1974mb}. Significant progress has been made in two-dimensional and supersymmetric CFTs, and a variety of methodologies, such as the large quantum number expansion  \cite{Hellerman:2015nra, Komargodski:2012ek}, conformal bootstrap \cite{Rattazzi:2008pe}, as well as numerical simulations, have been developed to tackle different dynamical sectors of a wide class of CFTs. One major hurdle is constituted by the evaluation of scaling dimensions of generic neutral composite operators built out of several fields. In fact, these operators are notoriously hard to access even with the conformal bootstrap approach, which works best for the low-lying spectrum. Even in weakly-coupled theories, perturbation theory breaks down when determining correlators of these operators \cite{Goldberg:1990qk, Badel:2019oxl}. Additionally, operator mixing under renormalization group (RG) flow leads to anomalous dimension matrices whose size increases rapidly with the number of fields constituting the composite operators. This fact hampers the determination of the physical spectrum. Neutral composite operators are also relevant beyond CFTs as they appear in effective field theories such as the one employed to investigate new physics beyond the Standard Model  (SMEFT) \cite{Buchmuller:1985jz, Grzadkowski:2010es, Brivio:2017vri}  \footnote{We refer the reader interested in the multi-loop renormalization of the SMEFT to \cite{Jenkins:2013zja, Jenkins:2013wua, Alonso:2013hga} for the complete $1$-loop renormalization and to \cite{Bern:2020ikv, Aebischer:2022anv, Jenkins:2023bls, DiNoi:2024ajj, Duhr:2025zqw, Born:2024mgz} for recent $2$-loop results.}. General properties of the anomalous dimension matrix for a wide class of theories have been explored in e.g., \cite{Craigie:1983fb, Alonso:2014rga, Elias-Miro:2014eia, Cheung:2015aba, Craig:2019wmo, Bern:2019wie, Jiang:2020rwz, Cao:2021cdt, Cao:2023adc}.

Recently, in \cite{Antipin:2024ekk,Antipin:2025ilv}, we put forward a semiclassical framework that allows us to determine the spectrum of scaling dimensions associated with families of composite neutral operators, made out of $n$ scalar fields, transforming in the traceless symmetric Lorentz representations, in the double-scaling limit 
\be \label{dsl}
n\rightarrow \infty \,, \qquad \lambda \rightarrow 0 \,, \qquad  \lambda n \ \text{fixed}  \,, 
\ee
where $\lambda$ collectively encapsulates the ensemble of couplings of the theory. 

In this work, we lay the foundations of the approach and pedagogically develop all the necessary steps, enabling the reader to  grasp the framework and readily apply it beyond the physically relevant examples presented here. To this end we start by considering a generic CFT in $d>2$ dimensions with a continuous global symmetry group $\mathcal{G}$ and a scalar sector with fields $\phi_A$, where $A$ is a collective index for a given representation of $\mathcal{G}$. Here we consider spin $s$ composite operators in the singlet representation of $\mathcal{G}$ with $n$ fields and transforming according to the traceless symmetric representations of the Lorentz group $SO(d-1,1)$, which can be labeled via the positive integer $s$. We generically denote their scaling dimensions as $\Delta_{n,q_\ell}$, where $q_\ell$ denotes a given set of integers specifying the operator and whose meaning will become clear later on. Following the path set in \cite{Antipin:2024ekk, Antipin:2025ilv}, the desired conformal dimensions map, via the state-operator correspondence, into the energy spectrum  $E_{n,q_\ell}$ of states obtained by quantizing homogeneous time-periodic solutions on the cylinder $\mathbb{R}\times S^{d-1}$ with $r$ the radius of the sphere. Interestingly, for the $\phi^4$ model considered in \cite{Antipin:2025ilv}, these periodic orbits develop an instability at large values of $\lambda n$.   Unstable periodic solutions of the equation of motion, that, unlike generic trajectories in chaotic systems, do not span all of the phase space, are known as \emph{classical scars} \cite{Heller:1984zz}. The name reflects the observation that, despite being atypical trajectories of the system, they leave a strong imprint, i.e., a scar, on the wavefunction since several energy eigenstates localize around them \cite{Heller:1984zz, Berry, Bogomolny}. 

In the limit \eqref{dsl} one can perform a semiclassical expansion in the inverse of the number of fields $n$. This leads to the following schematic form for the scaling exponents: 
\begin{equation} \label{dsl1}
    \Delta_{n,q_\ell} = r E_{n,q_\ell} = n \sum_{i=0} \frac{C_i(\lambda n,q_\ell)}{n^i} \ ,
\end{equation} 
where $C_0$ has been first computed in \cite{Antipin:2024ekk} for the critical $\phi^3$,  $\phi^4$, and $\phi^6$ theories. There, it was shown that in the small $\lambda n$ limit we recover the known perturbative results while predicting an infinite series of higher order terms. To this order in semiclassics the results depend purely on $n$, implying that the full spectrum of operators is degenerate.  We further unveiled that the generic large $ \lambda n$ behavior assumes the form $\Delta_{n, q_\ell} \propto n^{d/(d-1)}$. The  $C_1(\lambda n,q_\ell)$  term was presented first  in \cite{Antipin:2025ilv}, for the Ising $\phi^4$ CFT.  In the same work, within the semiclassical expansion, the   dependence on $q_\ell$ first appeared showing how the spectrum degeneracy is lifted at NLO. One can view these results as a novel quantum field theory (QFT) application of the Gutzwiller trace formula \cite{Gutzwiller:1971fy}, which expresses the energy spectrum as a trace over the periodic orbits and plays a key role in the semiclassical understanding of quantum chaos. 

We benchmark and elucidate our  framework by considering the $O(N)$-invariant $\phi^4$ theory and $\phi^6$ in $d=4-\epsilon$ and $d=3-\epsilon$ dimensions, respectively. By varying $N$, the critical $\phi^4$ theory falls into different universality classes describing continuous phase transitions spanning a rich variety of real-world three-dimensional condensed matter systems such as dilute polymer solutions (self-avoiding walks) $N=0$, superfluid helium for $N = 2$, and isotropic magnets for $N = 3$, to name a few (see \cite{Pelissetto:2000ek} for a comprehensive review of the physical applications of the $O(N)$ CFT). The $N=1$ theory describes the Ising universality class.   Complementary methodologies have been devised to extract relevant information such as the  state-of-the-art  $\epsilon$-expansion \cite{Henriksson:2025hwi, Henriksson:2025vyi, Schnetz:2022nsc, Kompaniets:2017yct} which provides remarkably good predictions for some of the critical exponents relevant here, with some of the most accurate results arising from numerical conformal bootstrap in $d=3$ \cite{Simmons-Duffin:2016wlq, Kos:2013tga}. Last but not least, the four-dimensional $O(4)$ $\phi^4$ theory is the Higgs sector of the Standard Model. 

The $O(N)$ $\phi^6$ theory in $d=3-\epsilon$ dimension provides a description of tricritical points, where multiple phase transition lines meet, of $O(N)$ invariant physical systems with two tunable parameters (e.g., both chemical potential and temperature). For $N=1$ one has the tricritical Ising CFT, which has been thoroughly investigated in the past. For general $N$, the scaling dimensions of the $\phi^n\,, \ n=1,2,4,6$ operators, needed to renormalize the model, are currently known to the $6$-loop order in the $\epsilon$ expansion \cite{Bednyakov:2025usv, Hager:2002uq}. An interesting property of this theory is that the one-loop beta function of the self-coupling vanishes. In turn, this property trivializes the renormalization of the next-to-leading semiclassical contribution, thereby providing a neat scenario to illustrate the general framework.

We verify our results via explicit comparison with perturbative computations in the small $\lambda n $ limit while yielding a new infinite series of predictions for higher order terms, substantially increasing our knowledge of CFT data. Furthermore, when expressed in terms of the couplings, our results can be extended away from conformality where they map directly into the eigenvalues of the anomalous dimension matrix, thereby fully solving the involved mixing problem in the neutral composite operator sectors of scalar QFTs. 

We organize our work as follows. In section \ref{saditappo} we start elucidating the framework by discussing the free field theory limit and then turning to the general interacting case in Sec.~\ref{canovaccio}. In Sec.~\ref{energypath}, we review the semiclassical quantization of time-periodic solutions in QFT developed in \cite{Gutzwiller:1971fy, Dashen:1974ci, dhn} which is based on the Floquet/Bloch theory \cite{floquet}. The main result in this section will be a formula for the $1$-loop energy spectrum $E_{n,q_\ell}$ expressed via a sum over the \emph{stability angles}  characterizing the stability of the classical periodic orbit.  Next, in Sec.~\ref{phi4} we apply the framework to the $O(N)$ model with $\phi^4$ interactions in $d=4-\epsilon$ dimensions. Here we first determine the scaling dimensions to the next leading order and then provide an algorithm to identify the primaries of the theory for any given set of integers $q_\ell$.  A similar analysis is performed for the $\phi^6$ theory in $d=3-\epsilon$ dimensions in  Sec.~\ref{phi6}. We offer our conclusions and suggest future directions in Sec.~\ref{outlook}. We include a series of appendices in which we provide further details of our computations.

\section{Free field theory} \label{saditappo}

It is instructive to start our journey from the free field theory case, reviewing the discussion in \cite{Cuomo:2024fuy}. To this end, we consider a non-interacting field theory of a single scalar field $\phi$ in $d$-dimensions and the two-point function of the $\phi^n$ operators
\begin{equation} 
\langle \phi^n(x_f) \phi^n(x_i) \rangle \ = \int \mathcal{D} \phi  \ \phi^n(x_f)\phi^n(x_i) e^{- \int d^d x \frac{1}{2}(\partial \phi)^2} \,.
\end{equation} 
After exponentiating the operator insertions, the above correlator can be determined via a semiclassical expansion dominated by the extrema of the following action
\begin{equation} \label{Seff}
   \mathcal{S}_n=   \int d^d x \left(  \frac{1}{2}(\partial \phi)^2 \right) - n \, \left(\log  \phi (x_f) + \log  \phi (x_i) \right) \ .
\end{equation}
The solution of the equation of motion (EOM) stemming from the action \eqref{Seff} yields $\phi_\text{cl} \equiv v(x)$, which can be expressed in terms of the Green function $G(x)$
\begin{align} \label{green}
v (x)&=\int d^dy  \ G(x-y)\frac{n}{v (y)}\left(\delta ^d\left(y-x_f\right)+\delta ^d\left(y-x_i\right)\right)=\frac{n G\left(x-x_f\right)}{v \left(x_f\right)}+\frac{n G\left(x-x_i\right)}{v \left(x_i\right)} \,,   
\end{align}
with
\be
G(x) =\frac{1}{(d-2) \Omega_{d-1}  \abs{x}^{d-2}} \ \,,
\ee
and
\be
v\left(x_f\right) v \left(x_i\right)=n G\left(x_f-x_i\right) \,,
\ee
where the divergence appearing when taking the   $\lim_{x\rightarrow 0}G(x)$ is regularized to zero. 
 By replacing the solution in the action, we recover the standard expression for the correlator
\begin{equation} \label{correlatorn}
\langle \phi^n(x_f) \phi^n(x_i) \rangle \simeq
e^{n(\log n-1)} [G(x_f-x_i)]^n\,. 
\end{equation}
The solution \eqref{green} assumes a  simple form after we map it onto a cylinder $\mathbb{R}\times S^{d-1}$ with radius $r$ and further use conformal symmetry to move the operator insertions at $t_{f/i}=\pm i \infty$, in the Lorentzian cylinder time coordinate $t$ 
\begin{equation}\label{eq_phi_cyl}
v(t)=2\sqrt{\frac{n}{(d-2)\Omega_{d-1} r^{d-2} }} \cos( \mu t+t_0)\,, \qquad \Omega_{d-1} = \frac{2 \pi ^{d/2}}{\Gamma \left(\frac{d} {2}\right)} ,
\end{equation}
where $t_0$ is a zero-mode of the solution stemming from time-translation symmetry (the free parameters in Eq.~\eqref{green}) and
\be \label{confmass}
\mu =\frac{d-2}{2r} \,,
\ee 
is the mass arising from the conformal coupling of the scalar to the Ricci curvature of the sphere. Naturally, $v(t)$ in \eqref{eq_phi_cyl} is also  a periodic homogeneous solution of the classical equation of motion on the  cylinder 
\be
\frac{d^2v}{dt^2} +\mu^2 v  = 0 \ .
\ee
According to Cardy's state-operator correspondence, a given scaling dimension $\Delta_{n,q_\ell}$ becomes the energy $E_{n,q_\ell}$  of the corresponding state on the cylinder, i.e.
 \be
  \Delta_{n,q_\ell} = r E_{n,q_\ell}  \ .
\ee
Therefore, from Eq.~\eqref{correlatorn} we learn that Eq.~\eqref{eq_phi_cyl} describes a primary state on the cylinder with energy $\Delta_{n,q_\ell}=\left(\frac{d-2}{2} \right) \,  n$ and is a periodic homogeneous solution of the EOM on the cylinder supplemented by the Bohr-Sommerfeld quantization condition
\begin{equation} \label{BS}
I  = 2\pi n \ , 
\end{equation}
where the action variable $I$ reads
\be\label{eq:actionvar}
 I = \oint \Pi d \phi =  \Omega_{d-1} r^{d-1} \int_{0}^{{\cal T}} \left(\frac{dv}{dt} \right)^2 \,dt \,, 
\ee
with $\Pi =\Omega_{d-1} r^{d-1}\dot \phi$  the conjugate momentum and $\T =2\pi/\mu$ the oscillation period. In Section~\ref{energypath}, we will show how the Bohr-Sommerfeld condition naturally arises when quantizing time-periodic field solutions from a path integral perspective.

\section{The canovaccio as semiclassical blueprint} \label{canovaccio}

We now start addressing the interacting dynamics. Consider a perturbatively renormalizable weakly-coupled quantum field theory with upper critical dimension $d$ and internal global symmetry group $\mathcal{G}$. We momentarily assume $\mathcal{G}$ to be continuous and that the scalar sector of the theory is made of scalar fields $\phi_{A}$, where $A$ is a collective index specifying a given irreducible representation of $\mathcal{G}$. At the end of the section, we will comment on discrete symmetries and the presence of scalar fields transforming in multiple representations of $\mathcal{G}$. We collectively denote the coupling constants of the theory as $\{\kappa_i\}$. One can engineer a perturbative Wilson-Fisher fixed point by tuning all the mass terms to zero and moving from $d$ to $d-\epsilon$ dimensions, with $\epsilon \ll 1$, where one can find (complex) zeros of the $1$-loop beta functions $\{\kappa_i\}= \{\kappa_i^*(\epsilon)\}$. We assume that the so-constructed theory is conformally invariant while generically being non-unitary.

In the free field theory limit defined by $\epsilon \to 0$ and $\{\kappa_i^*(\epsilon) \}\to 0$, we can build a generic primary operator from components of the field and derivatives. Hence, a primary operator in the singlet representation of $\mathcal{G}$ can be schematically denoted as 
\begin{align}\label{eq: composites}
   \cO_{s,p,n,i}=\partial^s \Box^p \phi^n \ . 
\end{align}
With this notation, we intend the appropriate linear combination of terms with $n$ fields and $2p$ derivatives with fully contracted $\mathcal{G}$ and Lorentz indices and $s$ uncontracted derivatives such that $\cO_{s,p,n,i}$ is a conformal primary. The additional index $i$ distinguishes between different primaries with the same schematic form, i.e., the same values of $s$, $p$, and $n$.
The introduction of the interaction substantially affects the spectrum of singlet composite operators via operator mixing. That is, bare operators of the form of Eq. \eqref{eq: composites} mix under renormalization. Henceforth, a primary operator is no longer a single operator of the form of Eq. \eqref{eq: composites}, but a linear combination of renormalized operators that diagonalize the anomalous dimension matrix. We here follow the standard convention of schematically denoting the interacting conformal primaries by those to which they reduce in the corresponding free field theory limit in $d$ dimensions \footnote{The underlying assumption is that, for every local operator for the free field theory obtained by setting $\kappa_i = 0 \ \forall \ i$ in $d$ dimensions, there exists a local operator of the interacting CFT in $d-\epsilon$ dimensions, which reduces to the former in the $\epsilon \to 0$ limit.}. In what follows, we will focus on primary operators transforming according to the traceless symmetric representation of the Lorentz group, which can be labeled by $s$. As mentioned in the introduction, we denote their scaling dimensions as $\Delta_{n,q_\ell}$, where the precise meaning of this notation will be explained in Sec.~\ref{energypath}. In the perturbative $\epsilon$-expansion $\Delta_{n,q_\ell}$ takes the form
\be \label{roperto}
 \Delta_{n,q_\ell} \sim \sum_{k=0}^\infty P_{k}(n,q_\ell)\epsilon^k \ .
\ee
We further assume that $P_{k}(n,q_\ell)$ is analytic around infinite $n$, such that it can be formally expanded as
\begin{equation} \label{cuoppo}
   P_k(n,q_\ell) \sim n^{\alpha_k}\sum_{i=0}^\infty c_{ik}(q_\ell) \ n^{-i} \,.
\end{equation}
The overall power $\alpha_k$ is model-dependent.  For instance, in the $\phi^4$ theory, one has $\alpha_k=k+1$ as it will be deduced either via the semiclassical approach or via ordinary diagrammatic arguments \cite{Badel:2019oxl}. The generic form of the $1/n$ expansion for $P_k$ depends further on a given operator and the specific loop order $k$. For example, the tree-level coefficient $P_0$ is always linear in $n$, i.e., $c_{i0} = 0 \ \forall \ i\ge 2$. For some operators $P_k$ is a polynomial for any value of $k$, and so the sum over $i$ in Eq.~\eqref{cuoppo} terminates for $i=\alpha_k$. Nevertheless, operator mixing typically leads to a more involved $n$ dependence. 

Our general strategy for the semiclassical evaluation of $\Delta_{n,q_\ell}$ consists of starting from homogeneous time-periodic classical solutions of the interacting theory on the cylinder. As long as the interaction strength is small, these periodic configurations can be seen as a perturbative deformation of the free field theory solution \eqref{eq_phi_cyl}, and, therefore, we can expect them to be still related, via operator-state correspondence, to a primary operator of the CFT. We can then semiclassically compute the corresponding energy spectrum $E_{n,q_\ell} = r \Delta_{n,q_\ell}$ and validate the approach via explicit comparison with the $\epsilon$-expansion. On general grounds, these solutions assume an instantonic-like form $v(t) \sim {\epsilon^{-\beta}}\,\tilde{v}(t, m)$, for some positive $\beta$, where $m$ is a parameter to be fixed via the Bohr-Sommerfeld condition in Eq.~\eqref{BS} \footnote{While this form of the solution is correct for the simple examples considered in this work, multiple parameters may appear in more involved cases. Nevertheless, no changes in our scaling argument are expected.}. The corresponding saddle point action scales as $\epsilon^{-2 \beta}$, which is naturally the loop counting parameter of the semiclassical expansion. Moreover, since the action variable $I$ is quadratic in the fields, the Bohr-Sommerfeld condition can be seen as an equation determining $m=m(\epsilon^{2\beta} n)$. We are then led to consider the double scaling limit 
\be \label{dsleps}
 \epsilon \to 0 \,, \quad n \to \infty \,, \quad \epsilon n^{\frac{1}{2\beta}} = \text{fixed.}
\ee
Accordingly, the semiclassical expansion is ultimately controlled by inverse powers of $n$ where, as mentioned, the latter should be interpreted as the number of fields entering in the definition of the operator in the free field theory limit. Consequently, the semiclassical expansion of $\Delta_{n,q_\ell}$ takes the form anticipated in Eq.~\eqref{dsl1}
\be \label{dslOK}
 \Delta_{n,q_\ell} \sim n \sum_{i=0}^{\infty} \frac{C_i \left(\epsilon n^{\frac{1}{2\beta}},q_\ell \right)}{n^i} \,.
\ee
By comparing the above to Eq.~\eqref{roperto}, we deduce that the perturbative expansion of the $C_i$ coefficients reads
\be \label{comparis}
C_i \sim \sum_{k=0}^\infty c_{ik} \ \left(\epsilon n^{\frac{1}{2\beta}} \right)^{k} \,.
\ee
Therefore, the coefficients $c_{ik}$ can be read off from the small $\epsilon n^{\frac{1}{2\beta}}$ expansion of $C_i$, providing a nontrivial test of the approach. Note that the above implies $\alpha_k = \frac{k}{2 \beta} + 1$. This relation can be traced back to having started with a perturbatively renormalizable theory where the form of the interactions is dictated by naive power-counting arguments. For instance, in the $\phi^4$ theory, the structure of the EOM implies $\beta=1/2$. The leading term $C_0$ of the semiclassical expansion is the classical energy of the periodic orbit on the cylinder and, as evident from Eq.~\eqref{comparis}, resums the leading powers of $n$ at any loop order. Analogously, $C_1$, which can be obtained by considering the corresponding fluctuation functional determinant, resums the subleading powers of $n$ and so on. 

If the considered Wilson Fisher fixed point occurs at real values of the coupling, the approach provides a determination of the scaling dimension spectrum in the double scaling limit \eqref{dsleps}. One may then eventually consider the limit $\epsilon \to 1$ to obtain results for the physical CFT in $d-1$ dimensions. Of course, this is a rough  extrapolation akin to the one performed within the standard  $\epsilon$-expansion.  Interestingly, in the case of the Ising $\phi^4$ theory, the NLO semiclassical expansion extrapolated to $\epsilon \to 1$ and $n \sim \order{1}$ provides results closer to the conformal bootstrap estimate than the NLO (i.e., $1$-loop) $\epsilon$-expansion already for $n=5$, and for $n \gtrsim12$ is expected to supersede the state-of-the-art $5$-loop $\epsilon$-expansion \cite{Antipin:2025ilv}.

Additionally, when rewritten in terms of the couplings $\{\kappa_i\}$, the small $\epsilon n^{\frac{1}{2\beta}}$ expansion of the $C_i$ coefficients reproduces the results obtained via ordinary perturbation theory for the associated anomalous dimensions  $\gamma_{n, q_\ell} $ for \emph{arbitrary values of the couplings} in the associated non-conformal $d$-dimensional theory. At the fixed point the $\gamma_{n, q_\ell} $ and $\Delta_{n, q_\ell} $ are related as
\be \label{anomalia}
 \Delta_{n, q_\ell} = \left( \frac{d-2}{2}\right) n + s + 2 p + \gamma_{n, q_\ell} \ .
\ee
It follows that the procedure of tuning the theory to a Wilson-Fisher fixed point can also be seen as a trick to employ the power of conformal symmetry even away from conformality. As first elucidated in \cite{Antipin:2020rdw}, for these purposes,  the Wilson-Fisher fixed point can either be achieved for real or complex values of the couplings, and therefore it does not matter whether the resulting CFT is unitary or not. Note that for the anomalous dimension of operators in nontrivial representations of $\mathcal{G}$, this procedure has been applied to several QFTs including the Standard Model \cite{Antipin:2023tar}. It is interesting to observe that, while the anomalous dimensions are scheme-dependent quantities, the terms with the leading power of $n$ at any loop order are not, since they all stem from the classical calculation of $C_0$. Analogously, the terms with the subleading powers of $n$ coming from the one-loop semiclassical term $C_1$ are scheme independent as long as the one-loop renormalization constants are scheme independent as well.

\vskip 1em

The following remarks are in order:

\begin{itemize}
    \item While we here focus on operators that are singlets of all the continuous symmetries of the theory, they can still transform in a nontrivial representation of a discrete symmetry group. For instance, in $Z_2$ invariant theories, a single periodic homogeneous saddle point spontaneously breaking $Z_2$ describes both $Z_2$-even and $Z_2$-odd operators \cite{Antipin:2025ilv}. However, the presence of a more involved discrete symmetry group may lead to the presence of multiple saddle points with different symmetry properties. Multiple saddle points generically emerge also when the scalar sector of the QFT contains scalar fields transforming according to different representations of $\mathcal{G}$.  

\item Even if our discussion revolves around singlet operators, the framework presented in this work actually encompasses the semiclassical approach for the calculation of anomalous dimensions of large charge operators transforming according to nontrivial symmetric representations of $\mathcal{G}$ in weakly coupled field theories \cite{Badel:2019oxl, Badel:2019khk, Antipin:2020abu}. The latter sub-case exhibits considerable simplifications as the time dependence of the saddle-point solution trivializes. Ultimately, the additional complications arising in the singlet sector mirror the ones stemming from operator mixing, and it is readily understood that their common conceptual origin lies in the absence of selection rules.

\item Operators transforming according to (partly) antisymmetric representations of the Lorentz and/or global symmetry groups can also be described semiclassically by considering a spatially inhomogeneous saddle point on the cylinder \cite{Hellerman:2017efx, Hellerman:2018sjf}. The latter also emerges in the description of correlation functions of primary operators with large spin \cite{Cuomo:2022kio, Choi:2025tql}. As it will become clear in the next section, the semiclassical framework presented here applies only when the spin of the operator is  parametrically smaller than $n$, i.e., for $s \ll n$.

\item One may consider different scaling limits such as $N \to \infty$ with $n/N$ fixed in $O(N)$-invariant CFTs or $(N_f -N_f^{\rm crit}) \to 0$ with fixed $(N_f -N_f^{\rm crit}) n$ for fixed points of the Banks–Zaks type. For large charge operators, these directions have been explored in e.g., \cite{Alvarez-Gaume:2019biu, Antipin:2022naw}.

\item A relevant application of the approach is in deducing further novel perturbative results by merging semiclassical and ordinary diagrammatic computations. In fact, in many cases, the $k$-th perturbative order $P_k(n,q_\ell)$ is a polynomial of degree $\alpha_k$ in $n$. Then the NLO semiclassical calculation provides two of the coefficients of the polynomial while the remaining ones can be further determined by matching to available $k$-loop perturbative results for $\alpha_k-2$ fixed values of $n$. This strategy has been previously employed in various CFTs \cite{Badel:2019oxl, Antipin:2020abu, Antipin:2022naw, Antipin:2023tar, Bednyakov:2023iuj} for charged operators and recently \cite{Antipin:2025ilv, Henriksson:2025vyi} for neutral operators in the Ising model.

\end{itemize}

Having outlined the general framework, in the next section, we will review the semiclassical quantization of homogeneous periodic solutions in quantum field theory.

\section{Semiclassical quantization of periodic solutions in QFT} \label{energypath}

To illustrate the program, let us start from a field theory describing a single real scalar field $\phi$ with action $\mathcal{S}$ and consider the trace over all the states of the theory for the propagator
\be \label{mellowyellow}
G(E) = \Tr \ \frac{1}{H-E} =\sum_n \frac{1}{E_n-E}  = i \int_0^\infty d \T  \ e^{i E\,\T} \ \Tr \  e^{-i H \T} \ ,
\ee
where $H$ is the Hamiltonian of the system. As expected, the bound state energy spectrum $E_n$ corresponds to the poles of the propagator and the sum is short-hand notation for taking into account both the discrete and continuum spectrum of the theory. For a scalar field theory on the cylinder the compactification of space constraints the spectrum to be discrete. 
The evolution kernel, being traced over the states, can be written in terms of a periodic path integral with period $\T$  
\be
\Tr \  e^{-i H \T} = \int \mathcal{D} \phi \ e^{i \mathcal{S}} \ .
\ee
We can now evaluate the path integral in the semiclassical approximation. The dominant contribution stems from the paths close to the classical periodic solutions $v(t)$ of the field equation of motion \cite{Gutzwiller:1971fy}. By expanding around the latter, we obtain 
\be
\Tr \  e^{-i H \T} \approx \mathcal{N} e^{i \mathcal{S}_{\rm cl}} |\det \cO^{(2)} |^{-1/2} \,,
\ee
where $\mathcal{N}$ is an unimportant normalization factor, $\mathcal{S}_{\rm cl}$ is the classical action evaluated on the $v(t)$ trajectory, and $\cO^{(2)}$ is the fluctuation operator appearing in the quadratic Lagrangian when expanding around $v(t)$. Using the method of separation of variables and separating the zero modes of $\cO^{(2)}$, we can write its determinant as 
\begin{align}
   \text{det}'\cO^{(2)} = \prod_{\ell} \text{det}' \mathcal{O}^{(2)}_\ell \ ,
\end{align}
where the prime denotes that the determinant does not include eventual zero modes, and the operator $\cO^{(2)}$ takes the form
\begin{align}
    \cO_\ell^{(2)} = -\partial_t^2  + V(t) - \Lambda(\ell)  , \quad V(t+\mathcal{T}) = V(\T)\,.
\end{align}
In writing the above we separated the full potential into a periodic time-dependent part $V(t)$ and a time-independent part $\Lambda (\ell)$, which includes the eigenvalues coming from the spatial part of the Laplacian, which we label by the collective index $\ell$. For instance, in our case, $\ell$ labels the eigenvalues of the Laplacian on $S^{d-1}$ whose eigenfunctions are the spherical harmonics. The periodicity of $V(t)$ follows from evaluating $\cO^{(2)}$ on the classical periodic configuration. 

According to the Gel'fand-Yaglom method \cite{Gelfand1960, Dunne:2007rt} we can compute the determinant of $\cO^{(2)}$ by considering two independent solutions $ \xi_{\ell,1}$ and $ \xi_{\ell, 2}$ of the following equation
\begin{align}\label{eq: hills equation}
    \left[-\partial_t^2 + V(t) - \Lambda (\ell)\right] \xi_{\ell,i}(t) = 0 \,, \quad i=1,2 \ \,,
\end{align}
subject to the boundary conditions
\begin{align}
    \xi_{\ell,1}(0) = 1, \quad \xi'_{\ell,1}(0) = 0, \quad \xi_{\ell,2}(0) = 0, \quad \xi'_{\ell,2}(0) = 1 \,. 
\end{align}
Equation \eqref{eq: hills equation} is of the Hill's type \cite{floquet}. Hill's equation appears in the description of the periodic orbits of physical systems. The stability of a classical orbit can be described in terms of the so-called \textit{monodromy matrix} $\mathcal{M}$ whose eigenvalues remarkably encode the spectral information of $\cO^{(2)}$. Indeed, according to Floquet/Bloch theory \cite{floquet}, which provides the general formalism for describing the solutions of Hill's equation, $\xi_{1,\ell}$ and $\xi_{2,\ell}$ satisfy
\begin{align}
    \label{eq: area preservering map}
    \begin{pmatrix}
        \xi_{\ell,1}(t+\mathcal{T})\\
        \xi_{\ell,2}(t+\mathcal{T})
    \end{pmatrix} = \mathcal{M} \begin{pmatrix}
        \xi_{\ell,1}(t)\\
        \xi_{\ell,2}(t)
    \end{pmatrix}, \quad \mathcal{M}=\begin{pmatrix}
        \xi_{\ell,1}(\mathcal{T}) & \xi'_{\ell,1}(\mathcal{T})\\
        \xi_{\ell,2}(\mathcal{T}) & \xi'_{\ell,2}(\mathcal{T})
    \end{pmatrix} \ .
\end{align}
The two eigenvalues of $\mathcal{M}$ can be written as $e^{\pm i \nu_\ell}$ where  the $\nu_\ell$ variables are known as \emph{stability angles}. A given periodic orbit is stable if and only if the stability angles are real. In what follows, we will assume a stable orbit and postpone the discussion of instabilities to Sec.~\ref{Instambul}. While it is clear that only $\nu_\ell \ \text{mod} ~ 2 \pi$ is physical, we here adopt conventions according to which the stability angles associated with the zero modes of $\cO^{(2)} $ vanish. Remarkably, the determinant of $\cO^{(2)}$ admits a simple expression in terms of the stability angles. In fact, applying the Gel'fand-Yaglom method, we obtain \cite{Dunne:2007rt}
\begin{align}
    \text{det}' \cO_\ell^{(2)} = 2 -\xi_{\ell,1}(\T) - \xi^\prime_{\ell,2}(\T)  = \det(1-\mathcal{M}) = 4 \sin^2( \nu_\ell/2 ) \ .
\end{align}
The solutions $\xi_{\ell\pm}$ that diagonalize $\mathcal{M}$ are called \emph{Bloch solutions} and are two linearly independent combinations of $\xi_{\ell,1}$ and $\xi_{\ell,2}$ which can be written as a product of a periodic function $\chi_{\ell,\pm}(t)$ and a plane wave
\be
\xi_{\ell,\pm} (t)=\chi_{\ell,\pm} (t) e^{\pm i \, p_\ell \, t}\, \,, \qquad \xi_{\ell,\pm}(t+\T)=e^{\pm i \, p_\ell\, \T}\ \xi_{\ell,\pm}(t) \,,
\ee
where $p_\ell =  \nu_\ell/\T$ is called the ``\emph{quasi-momentum}". Therefore, considering that the stability angles occur in pairs with opposite signs and equal magnitude, we arrive at
\begin{align}
     \text{det}' \mathcal{O}^{(2)}=4  \prod_{\nu_\ell>0}  \sin^2( \nu_\ell/2 ) \ .
\end{align}

Henceforth, only the knowledge of non-vanishing positive stability angles is sufficient for our purpose.  Employing the identity
\be
- 4 \sin^2 \nu_\ell=\frac{\left(1-e^{-i \nu_\ell }\right)^2}{e^{-i \nu_\ell }} \ ,
\ee
we obtain \cite{dhn}
\be \label{finalformula}
\Tr \  e^{-i H \T} = \sum_{\text{periodic orbits}} e^{i \left(\mathcal{S}_{\rm cl} -\frac{1}{2}\sum_{\nu_\ell >0} \nu_\ell \right)} \Delta_1 \Delta_2 \ ,
\ee
where $\Delta_1$ and
\be
\Delta_2 = \prod_{\nu_\ell>0} (1- e^{-i \nu_\ell})^{-1} \ ,
\ee
include, respectively, the contribution of the zero modes and the excited states. $\Delta_1$ stems from the integration over the zero modes associated with the symmetries of the theory \cite{dhn} and will play no role in the following discussion. Each factor of $\Delta_2$ can be rewritten as
\be
(1- e^{-i \nu_\ell})^{-1} = \sum_{q=0} e^{-i q\nu_\ell } \ ,
\ee
where each term of the sum corresponds to a distinct energy level. In practice, the factor $\Delta_{2}$ can be represented as $ e^{-i \sum_{\nu_\ell>0} q_{\ell} \nu_{\ell}} $ for a set of integers $q_{\ell}$, which identify the given energy level \cite{dhn}. 
By substituting Eq.~\eqref{finalformula} into Eq.~\eqref{mellowyellow} and omitting unimportant prefactors, we obtain
\be
G(E) \propto \int_0^\infty d \T \ \Delta_1 \ e^{i \left( \mathcal{S}_{\rm cl}-  \sum_{\nu_\ell >0}\left(\frac{1}{2}+q_\ell \right) \nu_\ell + E\,\T \right) } \ .
\ee
We now evaluate the integral over $\T$ via a saddle-point approximation. The saddle lies at
\be \label{saddleE}
E = -\frac{d \mathcal{S}_{\rm cl}}{d \T} +  \sum_{\nu_\ell >0}\left(\frac{1}{2}+q_\ell \right) \frac{d\nu_\ell}{d \T} \ ,
\ee
which determines $\T_{\rm saddle}=\T_{\rm saddle}(E)$, thereby fixing $\T$ as the period of the classical orbit with the given energy $E$. Taking into account that one can traverse the basic periodic orbit $k$ times, one arrives at
\be \label{GdiE}
G(E) \propto  \sum_{k=1}^\infty e^{i k \left(I+ \sum_{\nu_\ell >0}\left(q_{\ell}+\frac{1}{2}\right)\left(\T\frac{d\nu_{\ell}}{d\T}-\nu_{\ell}\right) \right)} = \frac{e^{i\left(I+ \sum_{\nu_\ell >0}\left(q_{\ell}+\frac{1}{2}\right)\left(\T\frac{d\nu_{\ell}}{d\T}-\nu_{\ell}\right) \right)}}{1-e^{i\left(I+\sum_{\nu_\ell >0}\left(q_{\ell}+\frac{1}{2}\right)\left(\T\frac{d\nu_{\ell}}{d\T}-\nu_{\ell}\right) \right)}} \ ,
\ee
where we used that the action variable $I$ introduced in Eq.~\eqref{eq:actionvar}, which can be seen as a function of $E$, is the Legendre transform of the classical action, that is 
\be
I(E) = \mathcal{S}_{\rm cl} -  \T \frac{d \mathcal{S}_{\rm cl}}{d \T} \ .
\ee
Therefore, the spectrum consists of a discrete set of bound states $E_{n,q_\ell}$ given by the values of $E$ solving Eq.~\eqref{saddleE} and satisfying the following quantization condition
\be
I(E)+\sum_{\nu_\ell>0}\left(q_{\ell}+\frac{1}{2}\right)\left(\T\frac{d\nu_{\ell}}{d\T}-\nu_{\ell}\right) = 2\pi n  \,, \label{kkk}
\ee
for a positive integer $n$. When this condition is satisfied, the trace of the propagator has a pole as evident from Eq.~\eqref{GdiE}. The result Eq.~\eqref{GdiE} for the semiclassical approximation of $G(E)$ is an instance of the so-called \emph{Gutzwiller Trace Formula} \cite{Gutzwiller:1971fy} that remarkably combines all the contributions stemming from the saddle-point evaluation of the propagator, the trace operation, and the integral over $\T$ into a single geometrical factor, namely the determinant of the monodromy matrix $\mathcal{M}$. 

The terms involving the stability angles represent genuine $1$-loop corrections stemming from the functional determinant. In the classical limit, the above reduce to the usual WKB result
\be \label{WKB}
I(E_{\rm cl}) = 2 \pi n \,, \qquad E = E_{\rm cl} = -\frac{d \mathcal{S}_{\rm cl}}{d \T}  \ .
\ee
 with $E_{\rm cl}$ the classical energy of the orbit. To simplify our formulae, we separate classical and higher-order effects following \cite{Beccaria:2010zn}. First, we note that all the UV divergences are contained in the sum over the positive stability angles appearing in the exponent of Eq.~\eqref{finalformula}, which needs to be renormalized (see Sec.~\ref{renatino}). In particular, in a renormalizable QFT, it is always possible to make the combination $\mathcal{S}_{\rm cl} - \frac12 \sum_{\nu_\ell >0} \nu_\ell $ finite when expressed in terms of the renormalized parameters \cite{dhn}. Generically, denoting as $\tilde{\mathcal{S}}$ the renormalized action, the procedure will generate higher-order effects according to
\be \label{SoverT}
 -\frac{d \tilde{\mathcal{S}}}{d \T} = E(n) +\delta E_1 + \dots
\ee
Here $E(n)$ contains the classical energy $E_{\rm cl}$ plus subleading (in the semiclassical expansion) corrections stemming from the second term in the LHS of Eq. (\ref{kkk}), which can be viewed as a $1$-loop quantum correction to the Bohr-Sommerfeld condition \eqref{BS}. We further singled out a one-loop term $\delta E_1$ generated by the renormalization procedure and generically proportional to the $1$-loop coefficients of the beta functions of the couplings as illustrated in Sec.~\ref{renatino} for the $\phi^4$ theory. The dots denote higher-order semiclassical contributions.

Accordingly, we can write the energy of a quantum state as
\be 
E_{n,q_\ell} = E(n)+ \delta E_1 +  \sum_{\nu_\ell>0}\left(q_{\ell}+\frac{1}{2}\right) \frac{d\nu_{\ell}}{d\T}  \ .\label{fr}
\ee
All the terms on the RHS of Eq.~\eqref{fr} but the first one are automatically $1$-loop in the semiclassical expansion and, therefore, can be determined via the leading order Bohr-Sommerfeld condition in Eq.~\eqref{BS}. Expanding $E(n)$ as $E(n)=E_{cl}(n)+\delta E_2$ we can solve perturbatively Eq.~\eqref{kkk} to obtain  
\be
\delta E_2 =-\sum_{\nu_\ell>0}\left(q_{\ell}+\frac{1}{2}\right)\left(\frac{d\nu_{\ell}}{d\T}-\frac{\nu_{\ell}}{\T}\right) \ ,
\ee
where we used Eq.~\eqref{WKB} and that 
\be
\frac{\del I}{\del E}\rvert_{E=E_{\rm cl}} = \T  \,.
\ee
Therefore, keeping only terms up to NLO in the semiclassical expansion, we have
\begin{align} 
E_{n,q_\ell} &=  E_{cl}(n) +\delta E_1  + \delta E_2  + \sum_{\nu_\ell>0}\left(q_{\ell}+\frac{1}{2}\right) \frac{d\nu_{\ell}}{d\T}  \nonumber \\
&= E_{cl}(n)+   \delta E_1 +\frac{1}{\T}\sum_{\nu_\ell>0}\left(q_{\ell}+\frac{1}{2}\right) \nu_{\ell} \  .\label{pou}    
\end{align}
Since all the stability angles are positive, the ground state is attained for $q_\ell=0$. Finally, by invoking the state-operator correspondence, we arrive at 
\be \label{finafinal}
n C_0 = r E_{cl} \,, \qquad C_1 = r \delta E_1 + \frac{r}{\T}\sum_{\nu_\ell>0}\left(q_{\ell}+\frac{1}{2}\right) \nu_{\ell} \ .
\ee
This is the central result of this section. Let us comment on the realm of validity of this formula. First, we note that the manipulations performed to organize the semiclassical expansion, identifying the classical term and the quantum corrections, are legitimate only when $q_\ell \ll n \ \forall \ \ell$ \footnote{See \cite{Beccaria:2010zn} for how this condition arises in the equivalent Bohr-Sommerfeld-Maslov quantization framework.}. Moreover, if one considers an infinite number of nonzero $q_\ell$'s (e.g., taking $q_\ell = \ell$), additional UV divergences arise. As will become clear in the next section, $n$ and $\sum_{\nu_\ell>0} q_{\ell} \ell$ can be respectively interpreted as the number of fields and derivatives used in the construction of the operator corresponding to the given excited state, we conclude that Eq.~\eqref{finafinal} does not hold for operators whose number of derivatives is parametrically of the same order as the number of fields or larger. It is indeed expected that spatial inhomogeneities appear at large spin \cite{Cuomo:2022kio, Choi:2025tql}.

As evident from the formalism, the generalization to multiple fields involves the presence of a functional determinant for each field, which in turn leads to multiple sets of stability angles and associated occupation numbers $q_\ell$. This will indeed be the case for the $O(N)$ scalar theories investigated in the next sections.

We conclude the section by commenting on how the semiclassical calculation of the scaling dimension of large charge operators in weakly-coupled CFTs \cite{Badel:2019oxl, Badel:2019khk, Antipin:2023tar}, follows from the outlined formalism by considering the simple case where the time-dependence of the ground state is trivial and so Bloch solutions reduce to plane waves. As a consequence, the stability angles reduce to $\nu_\ell = \omega_\ell \T$, where $\omega_\ell$ denotes the dispersion relations of the quadratic fluctuations. Since the stability angles are linear in $\T$ the NLO corrections to the Bohr-Sommerfeld condition in Eq.~\eqref{kkk} vanish, and the latter reduces to the charge-fixing condition. At the same time, Eq.~\eqref{finafinal} reduces to the usual $1$-loop formula for the semiclassical energy as a sum over the single particle energies. It is worth adding that, while the classical solutions appearing in the study of correlators of large charge operators are typically stable, we expect the ones corresponding to neutral operators to become unstable, at least at sufficiently strong coupling and/or large $n$. In fact, in the former case, quantum scarring, i.e., the localization of energy eigenstates around periodic orbits, simply reduces to the Bohr correspondence principle in the presence of large conserved quantum numbers. 

Having outlined the general formalism, the rest of the paper is devoted to applying it to specific examples.

\section{The $O(N)$ $\phi^4$ theory in $d=4-\epsilon$} \label{phi4}

Equipped with the general semiclassical machinery developed in the previous sections, we proceed by applying it to the critical  $O(N)$ $\lambda(\phi_a \phi_a)^2$ theory in $d=4-\epsilon$ dimensions
\begin{equation}
    \mathcal{L} =\frac{1}{2}(\partial \phi_a)^2 - \frac{\lambda}{4} (\phi_a \phi_a)^2 \,, \qquad a=1, \dots, N \,,
\end{equation}
where the fixed point coupling occurs at
\be \label{chiodofisso}
\lambda_* =\frac{8 \pi ^2 \epsilon }{N+8} +  \frac{24 \pi ^2 (3 N+14) \epsilon ^2}{(N+8)^3} + \order{\epsilon^3} \ .
\ee
For later comparison, in App. \ref{add1loop} we review the perturbative results for the scaling dimension of various families of operators in the traceless-symmetric Lorentz representations (see also \cite{Henriksson:2022rnm} for a comprehensive review of the conformal data of the $O(N)$ CFT). 
\subsection{Classical solution and leading order: $C_0$}
We begin our analysis by reviewing the computation of the classical contribution  $C_0$ carried out in \cite{Antipin:2024ekk}. The Lagrangian on the cylinder reads
\begin{equation} \label{Lagcyl}
    \mathcal{L}^{(\text{cyl})} =\frac{1}{2}(\partial \phi_a)^2 -\frac{\mu ^2 }{2}  \phi_a \phi_a-\frac{\lambda}{4}( \phi_a \phi_a)^2 \ .
\end{equation}
Assuming a spatially homogeneous field configuration and a vanishing classical profile for $\phi_a \,, \ a=2, \dots N$, the time-dependent equation of motion for $\phi_{1,\text{cl}} \equiv v(t)$ takes the form of the quartic anharmonic oscillator
\begin{equation}
\frac{d^2v}{dt^2} +\mu^2 v + \lambda v^3 = 0 \ ,
\label{eom}
\end{equation}
whose solution reads \cite{Sanchez}
\begin{equation} \label{vevo}
v(t) = \, x_0 \, \cn(\omega t|m) \ ,
\end{equation}
where $\cn(\omega t|m) $ stands for the Jacobi elliptic cosine function. The initial amplitude $x_0$ and the frequency $\omega$ are given by 
\begin{equation}
    x_0 =\mu \sqrt{\frac{2 m}{\lambda (1-2 m)}} \,, \qquad  \omega = \frac{\mu}{\sqrt{1-2 m}} \,, \qquad 0 \leq m \leq 1/2 \,.
\end{equation}
The period of the solution is given by
\be
{\cal T} = 4 \KK(m)/\omega \,,
\ee
where $\KK(m)$ denotes the complete elliptic integral of the first kind. According to the discussion in Sec.~\ref{canovaccio}, the semiclassical expansion takes the form in Eq.~\eqref{dslOK} with $\beta = 1/2$. Accordingly, the classical energy of the solution yields $C_0$
\begin{equation}
\label{C0}
n C_{0} =r E_{\rm cl} =\Omega_{d-1}r^d\braket{T_{00}} \rvert_{d=4} =  \frac{2\pi^2  m \,(1-m) } {\lambda \,(1-2 m)^2} \ ,
\end{equation}
where $T_{\mu\nu}$ is the stress energy tensor and, because the classical theory is conformal, we can fix $d=4$.  The parameter $m$ is a function of $\lambda n$ obtained by inverting the Bohr-Sommerfeld condition \eqref{BS}, where
\begin{align}
\label{mlambdan}
    I = \frac{16 \pi^2  }{3  \lambda (1-2 m)^{3/2}} \left[ (2m-1)\ELE(m) + (1-m)\KK(m) \right] \ ,
\end{align}
and $\ELE(m)$ is the complete elliptic integral of the second kind. In the small $\lambda n$ limit, one can solve perturbatively Eq.~\eqref{BS} obtaining
\be \label{mla}
m=\frac{(\lambda n)}{2 \pi ^2}-\frac{21 (\lambda n)^2}{32 \pi ^4}+\frac{479 (\lambda n)^3}{512 \pi ^6} -\frac{22745 (\lambda  n)^4}{16384 \pi ^8} + \frac{1105887 (\lambda n)^5}{524288 \pi ^{10}}+ \dots \,.
\ee
By substituting the above into our expression for the classical energy Eq.~\eqref{C0} and tuning the coupling to its fixed point value Eq.~\eqref{chiodofisso}, we obtain
\begin{equation}
   n C_0 = n \left(1+\frac{1}{6}(\epsilon n)-\frac{17}{324}(\epsilon n) ^2+\frac{125}{3888}(\epsilon n) ^3-\frac{3563}{139968}(\epsilon n) ^4+\frac{29183}{1259712} (\epsilon n) ^5+\dots \right)\,, 
\end{equation}
in agreement with Eq.~\eqref{checkitout}. Notably, $C_0$ does not depend on $N$.
In the large $\lambda n$ limit, one has
\begin{equation} \label{largelambdanm}
    m = \frac{1}{2} -\pi  \left(\frac{\Gamma \left(\frac{1}{4}\right)}{6 \Gamma \left(\frac{3}{4}\right)}\right)^{2/3} \left(\frac{1}{\lambda n}\right)^{2/3}  + \mathcal{O}\left((\lambda n)^{-4/3}
\right) \,, 
\end{equation}
leading to 
\begin{equation} \label{Largeorder}
n C_0 =\left(\frac{3\Gamma \left(\frac{3}{4}\right)}{2^{5/4}\Gamma \left(\frac{1}{4}\right)}\right)^{4/3} \lambda^{1/3} n^{4/3} +4 \pi ^3 \left(\frac{ 6 \ \Gamma \left(\frac{3}{4}\right)}{\Gamma \left(\frac{1}{4}\right)^{7}} \right)^{2/3}\lambda^{-1/3}n^{2/3}  +  \mathcal{O}\left(n^{0} \right) \ .
\end{equation}

\subsection{Renormalization of the action} \label{renatino}
 
In this section, we differentiate between bare and renormalized couplings and denote the former by $\lambda_0$. Our goal is to renormalize the exponent of Eq.~\eqref{finalformula}, i.e., $\mathcal{S}_{\rm cl} -  \frac12\sum_{\nu_\ell >0} \nu_\ell$, employing dimensional regularization. As mentioned, when renormalizing the quartic coupling \footnote{Note that the fields do not get renormalized at the one-loop level.} in the classical action, one isolates the $1/\epsilon$ pole needed to cancel the one-loop UV divergences arising when summing over the stability angles. The bare classical action in $d=4-\epsilon$ reads
\be
\mathcal{S}_{\rm cl}(\lambda_0) =\Omega_{d-1} r^{d-1} \int_0^\T  \, dt \,\left( \frac{1}{2} \left( \frac{d v}{d t} \right)^2-\frac{\mu ^2 }{2} v ^2 -\frac{\lambda_0 }{4}  v ^4 \right) = -\frac{\pi ^{2-\frac{\epsilon }{2}} (\epsilon -2)^3 r^{-\epsilon } s(m)}{\lambda_0 \Gamma \left(2-\frac{\epsilon }{2}\right)} \,,
\ee
where
\be
s(m) =\frac{(m-1) (3 m-2) \KK(m)+(4 m-2) \ELE(m)}{3 (1-2 m)^{3/2}} \,.
\ee
At the one-loop order, the coupling is renormalized according to
\be
\lambda_0 = \lambda  M^{\epsilon } e^{\frac{\beta _0 \lambda }{\epsilon }}  \,, \qquad \beta_0= \frac{N+8}{8 \pi ^2} \ ,
\ee
where $\beta_0$ is the one-loop coefficient of the beta function and $M$ is an arbitrary RG scale. Therefore, we have
\be
\mathcal{S}_{\rm cl}(\lambda) =  - \frac{\pi ^{2-\frac{\epsilon }{2}} (\epsilon -2)^3 M^{-\epsilon } r^{-\epsilon } s(m)}{ \Gamma \left(2-\frac{\epsilon }{2}\right)}\left(\frac{1}{\lambda }-\frac{\beta _0}{\epsilon } + \order{\lambda}\right) \,.
\ee
The order $\order{\lambda^0}$ term is a quantum correction and in the $\epsilon \to 0$ limit it reduces to
\be \label{polepole}
  \frac{\pi ^{2-\frac{\epsilon }{2}} (\epsilon -2)^3 M^{-\epsilon } r^{-\epsilon }s(m)}{ \Gamma \left(2-\frac{\epsilon }{2} \right)}\frac{ \beta_0 }{\epsilon}= -8 \pi ^2 \beta_0 s(m) \left(\frac{1}{\epsilon }- \frac{1}{2}\left(2 + \gamma_E + 2\log \left(\sqrt{\pi}  M r\right)\right)+ \order{\epsilon}\right) \,,
\ee
where $\gamma_E$ denotes Euler's constant. As anticipated, the $1/\epsilon$ pole appearing in the RHS will cancel the UV divergences arising in the sum over the stability angles, such that the combination $\mathcal{S}_{\rm cl} - \frac {1} {2} \sum_{\nu_\ell > 0} \nu_\ell$ is finite in the $\epsilon \to 0$ limit. We denote as $\tilde{\mathcal{S}}$ the finite part of $\mathcal{S}_{\rm cl}$, which, neglecting higher-order contributions, reads
\be
\tilde{\mathcal{S}}(\lambda) =\pi^2 s(m)\left(-\frac{\pi ^{-\frac{\epsilon }{2}} (\epsilon -2)^3 M^{-\epsilon } r^{-\epsilon }}{\lambda \,  \Gamma \left(2-\frac{\epsilon }{2}\right)}+ 4 \beta_0 \left(2 + \gamma_E + 2\log \left(\sqrt{\pi}  M r\right) \right) + \order{\lambda} \right) \,.
\ee
By taking the derivative with respect to the period and evaluating the expression at the fixed point, we arrive at
\be
r\frac{\partial  \tilde{\mathcal{S}}(\lambda)}{\del \T} \Big \rvert_{\lambda = \lambda_*} =r\frac{\partial  \tilde{\mathcal{S}}(\lambda)}{\del m}\left(\frac{\del \T} {\del m}\right)^{-1} \Big \rvert_{\lambda = \lambda_*}= n C_0\left(\lambda _* n\right) -\frac{\lambda _*n}{2} \beta _0  C_0\left(\lambda _* n\right)  + \order{1/ n} \ ,
\ee
where we are keeping $\lambda_\ast n$ fixed. Substituting the above into Eq.~\eqref{saddleE} and comparing to \eqref{SoverT}, allows us to identify
\be \label{deltaE1}
r \delta E_1 = -\frac{\lambda _*n}{2} \beta_0 C_0\left(\lambda _* n\right) \ .
\ee
Finally, recalling that $m$ is a function of $\lambda_\ast n$, we have 
\be
I \Big \rvert_{\lambda = \lambda_*}= \tilde{\mathcal{S}} - \T \frac{\del \tilde{\mathcal{S}}}{\del \T} \Big \rvert_{\lambda = \lambda_*} = \frac{16 \pi ^2 ((1-m )\KK(m)+(2 m-1) \ELE(m))}{3 \lambda_*  (1-2 m)^{3/2}}  + \order{1/n} \,.
\ee
Therefore, we achieved our two-fold goal of computing $\delta E_1$ and isolating the pole in the renormalization of the action. 


 
\subsection{Fluctuation operators} \label{lameform}

\begin{figure}[t!]
\centering
\includegraphics[width=0.49\textwidth]{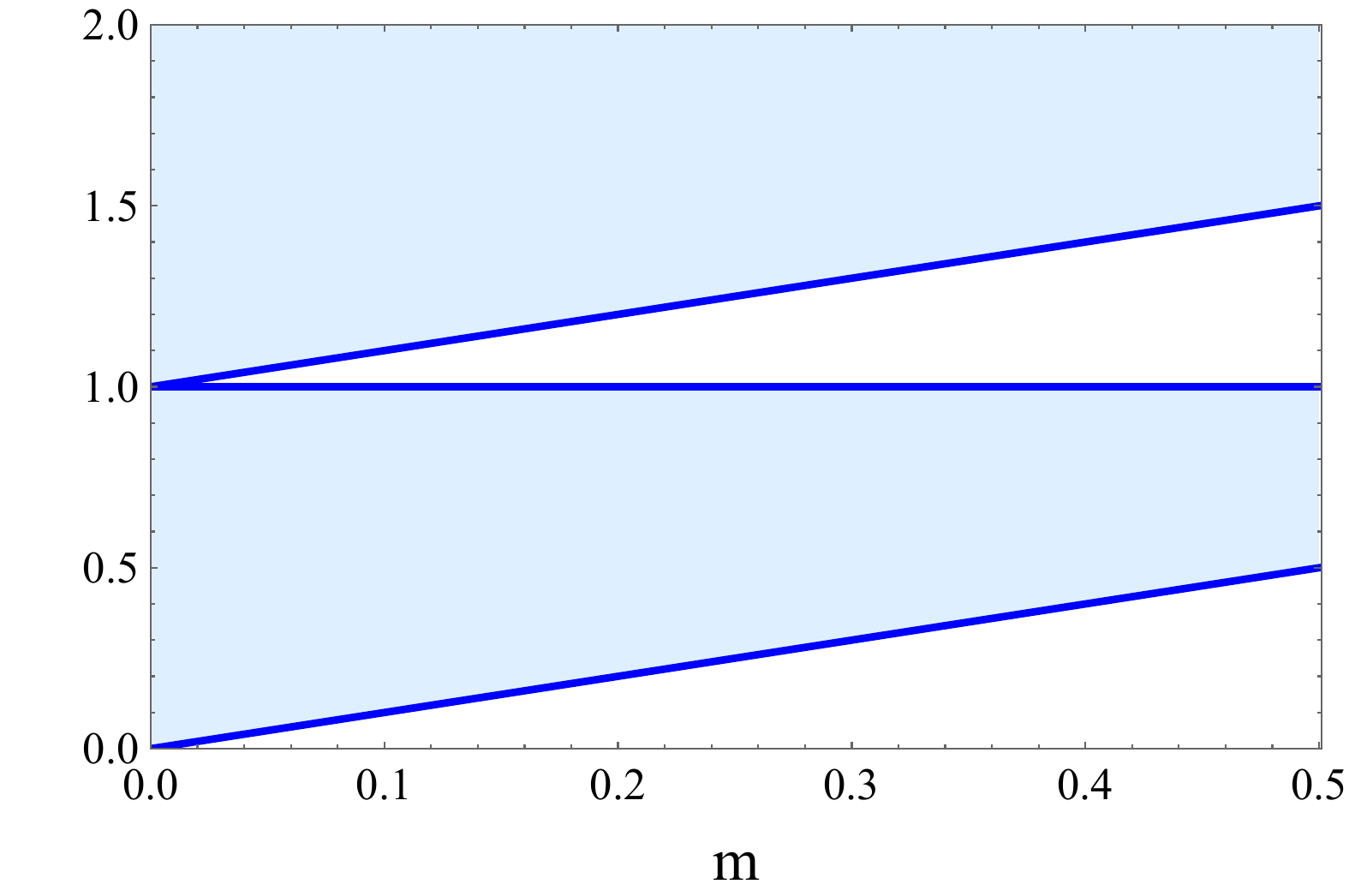} \includegraphics[width=0.49\textwidth]{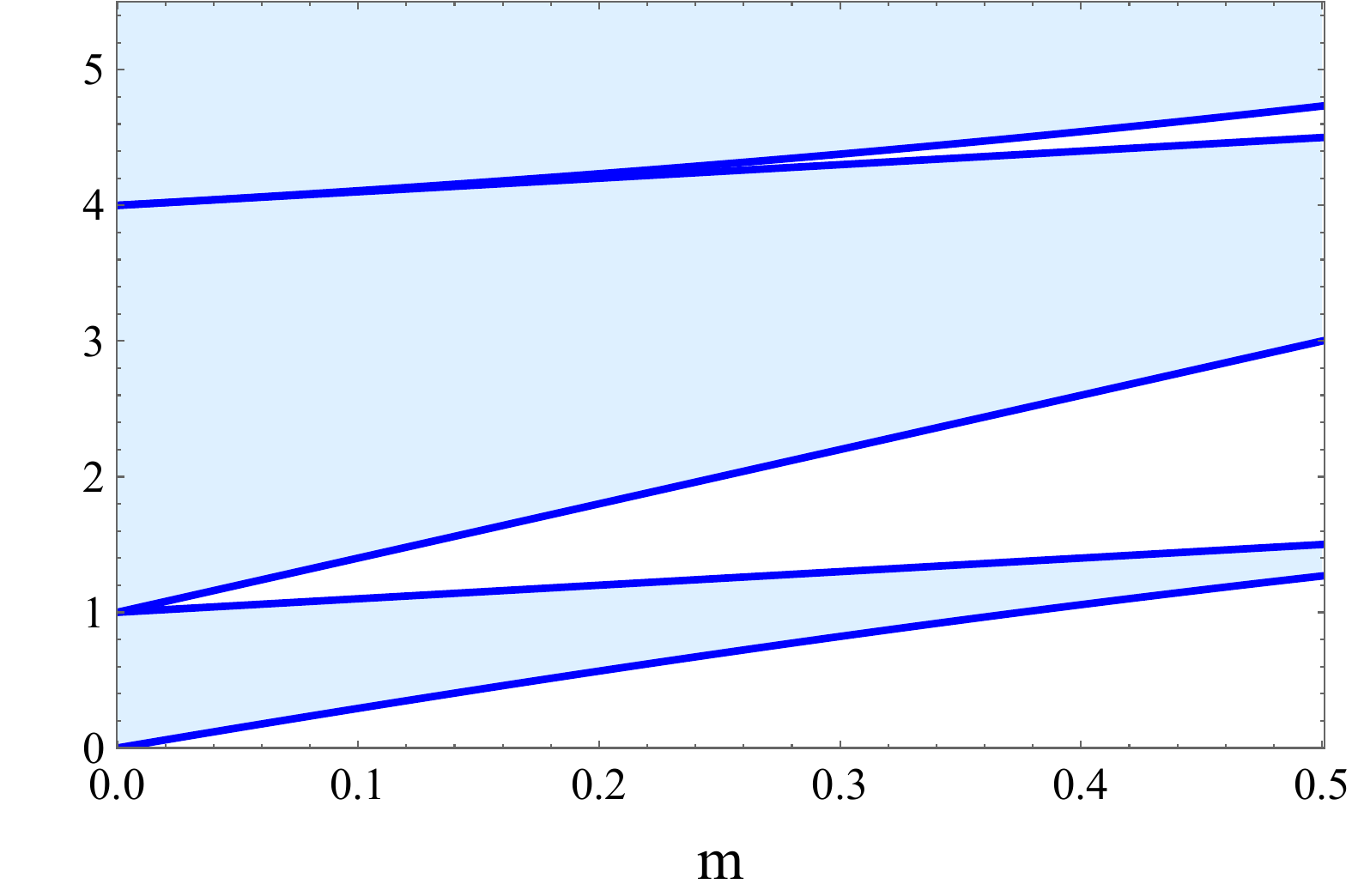} 
\caption{The band structure for the $\kappa =1$ (\emph{left}) and $\kappa =2$ (\emph{right}) Lam\'e operator $L_\kappa$ as a function of $m$. The allowed energy regions are filled in light blue. The solid lines denote the band edges whose values are given in the main text.} 
	\label{bandabassotti}
\end{figure}

In order to determine the stability angles of the fluctuation operators, we expand the action around the classical trajectory according to $\phi_1 = v(t) + \eta(\vec x,t)$ and $\phi_a = \tilde{\phi}_a(\vec x,t)$ for $a=2, \dots, N$ and obtain the following quadratic Lagrangian 
\be\label{eq:fluctuation4}
\mathcal{L}_2 =  \sum_{a=2}^N \frac{1}{2}\tilde{\phi}_a \cO_1\tilde{\phi}_a + \frac{1}{2}\eta \cO_2 \eta \ , 
\ee
where
\be
\cO_\kappa = -\partial _t^2+\Delta _S-\mu ^2-   \frac{\kappa (\kappa+1)}{2} \lambda \, v^2(t) \,, \qquad \kappa=1,2 \ ,
\ee
with $\Delta _S$ the Laplacian on $S^{d-1}$. By introducing $z= \frac{\mu t}{\sqrt{1-2 m}}$ we can rewrite both fluctuation operators as 
\be \label{rescaled}
\cO_{\kappa} = \frac{\mu^2 }{1-2m} L_{\kappa} \,, \qquad \kappa=1,2 \ ,
\ee
where $L_{\kappa}$ assume the form of the $\kappa$-gaps Lam\'e operator 
\be \label{eq: lame_operator}
L_\kappa = -\partial_z^2+\kappa (\kappa +1 ) \ m \ \sn(z|m)^2-\Lambda_\kappa(\ell) \ ,
\ee
with
\be \label{lambda12}
\Lambda_\kappa(\ell) = \kappa(\kappa+1) m +  (1-2 m)  A_\ell  \,, \qquad A_\ell \equiv \left(1+ \frac{J_\ell^2}{\mu^2} \right)= \left(1 + \frac{2 \ell}{d-2}\right)^2 \ .
\ee
$\Lambda_{\kappa}$ includes $\Delta_S$ in terms  of its  eigenvalues:
\be \label{EigJ}
J_\ell^2 =\frac{\ell (\ell + d -2)}{r^2}  \ .
\ee
The determinant of these operators is normalized by the free field theory determinant $\cO_0$, which can be written as
\be \label{freeearl}
\cO_0 = \frac{\mu^2}{1-2 m} L_0 \, .
\ee
It follows that the prefactor in Eq.~\eqref{rescaled} cancels in the ratio of determinants and does not play a role in the analysis. As discussed in Sec.~\ref{energypath}, we consider the equation 
\be\label{set}
L_\kappa \xi_{\ell,\pm}(z)= 0 \,,
\end{equation}
with periodic boundary conditions with rescaled period 
\begin{equation}
\frac{\T \sqrt{1-2m}}{\mu} = 4 \KK(m) \ .
\end{equation}
For $\kappa=1$, the band spectrum of the potential has  a single gap and two allowed bands, namely $\{m, 1\}$ and $ \{1 + m, \infty\} $ as illustrated in the left panel of Fig. \ref{bandabassotti}. 
For $\kappa=1$, the two independent Bloch solutions of \eqref{set} read \cite{book1}
\be\label{solslame}
\xi_{\ell,\pm}(z) = \frac{H(z\pm\alpha, \,|\,  m)}{\Theta(z \,|\,  m)}\,e^{\mp\,z\, Z(\alpha \,|\,  m)} \ ,
\ee
where $H, \Theta, Z$ respectively denote the Jacobi Eta, Theta, and Zeta functions reviewed in appendix \ref{jacobiapp}, and $ \alpha$ solves
\be\label{alphaeq}
{\sn}(\alpha\,|\,m) = \sqrt{\frac{1+m-\Lambda_1(\ell)}{m}}\,.
\ee
The periodic properties of the Jacobi functions imply the following expression of the stability angles for the operator with $\kappa = 1$
\begin{equation} \label{stab1}
    \nu_{1,\ell} =  - 4 i \KK(m)Z(\alpha\,|\,m) \,,
\end{equation}
where we used the fact that the stability angles are only defined $\text{mod} ~ 2 \pi$, and fixed the overall sign ($\pm$) to select the positive $\nu_{1,\ell}$. The index $\ell$ reminds us that the stability angles are labeled by $\ell$, which enters $\alpha$ through $\Lambda_1(\ell)$. 

The general Bloch solutions for $\kappa = 2$ are given by \cite{book1, Pawellek:2008st}
\begin{equation}\label{eq:GenSolution}
 \xi_{\ell,\pm}(z)=\frac{H(z\pm\alpha_+ \,|\,  m)H(z\pm\alpha_- \,|\,  m)}{\Theta(z\,|\,  m)^2}e^{\mp z(Z(\alpha_+\,|\,  m)+Z(\alpha_- \,|\,  m))} \,,
\end{equation}
where $\alpha_\pm$ solve 
\begin{align}
\label{eq:Translos}
 \sn^2(\alpha_\pm \,|\,  m)=\frac{4(1+m)-\Lambda_2(\ell)}{6m}\pm \frac{1}{2m}\sqrt{\frac{4}{3}(1-m+m^2)-\frac{1}{3}\left(\Lambda_2(\ell)-2(1+m)\right)^2} \,.
\end{align}
Therefore, the stability angles are
\be \label{stab2}
 \nu_{2,\ell} = 2 \pi - 4 i \KK(m) \left(\, Z(\alpha_+\,|\,m)+\, Z(\alpha_-\,|\,m) \right) \,.
\ee
In the $\kappa=2$ case, the spectrum has two gaps and three allowed bands, which are given by 
\be \label{2lamegap}
\left\{ 2 \left(m-\sqrt{1-m (1-m)}+1\right),  1+ m \right\}, \quad \left\{1+4m,  4+ m \right\}, \quad \left\{2 \left(m+\sqrt{1-m (1-m)}+1\right),  \infty \right\} \,.
\ee
The band structure is depicted in the right panel of Fig. \ref{bandabassotti}. A useful identity relating the Jacobi Zeta function and the complete elliptic integral of the third kind $\Pi$ is 
\be \label{useful}
Z\left({\sn}^{-1}\left(\frac{a}{\sqrt{m}}\,|\,m\right)\,|\,m \right)  = i\,\sqrt{1-a^{2}}
\,\sqrt{1-\frac{m}{a^{2}}}\left(1-\frac{\Pi(a^{2}\,|\,m)}{\KK(m)}\right) \,,
\ee
and allows us to evaluate the stability angles without having to solve equations \eqref{alphaeq} and \eqref{eq:Translos}.

For a general positive integer $\kappa$, beyond what is needed for this model (i.e., $\kappa=1,2$), the spectrum of the Lam\'e operator has $\kappa$ gaps and the Bloch solutions take the general form \cite{book1}
\be
\xi_{\ell,\pm}(z) = \prod_{j=1}^{\kappa}\frac{H(z\pm\alpha_j\,|\,  m) }{\Theta(z\,|\,  m)}e^{\mp z Z(\alpha_j\,|\,m)},
\ee
where the parameters $\alpha_j$ solve a set of $\kappa$ transcendental equations sometimes called Bethe ansatz equations of the Lam\'e potential \cite{Takemura_2003}. The associated stability angles are a sum of Jacobi Zeta functions 
\be
\nu_{\kappa,\ell} = \pm 4  i \KK(m) \sum_{j=1}^{\kappa} Z(\alpha_j\,|\,m)  \ .
\ee
In App.~\ref{smallnu} we discuss an alternative method for calculating the stability angles perturbatively in the small $\lambda n$ limit, which does not rely on knowing the exact Bloch solutions.

According to Eq.~\eqref{finafinal}, the leading quantum correction in the semiclassical expansion is given by
\be
C_1  = r \delta E_1 + \frac{r}{2 \T} \sum_{\ell=0}^\infty \left(n_\ell \left[ (N-1)  \nu_{1,\ell} +  \nu_{2,\ell} - N  \nu_{0,\ell} \right] +2 q_{1,\ell} \nu_{1,\ell} + 2 q_{2,\ell} \nu_{2,\ell} \right)\ ,
\ee
where we introduced two distinct sets of integers $q_{2,\ell}$ and $q_{1,\ell}$ associated with our two fluctuation operators and
\be
n_\ell = \frac{(2 \ell + d-2) \Gamma (\ell+d-2)}{\Gamma (d-1) \Gamma (\ell+1)} \ ,
\ee
 is the multiplicity of the eigenvalues of the Laplacian underlying the same degeneracy for the stability angles. Moreover, $\nu_{0,\ell}$ are the stability angles of the free field theory operator $\cO_0$ in Eq.~\eqref{freeearl}. Since the corresponding potential is static, the Bloch solutions are plane waves with quasi-momentum $\mu \sqrt{A_\ell}$ and  the stability angles read
\be
\nu_{0,\ell} = \mu \sqrt{A_\ell}  \T \ .
\ee
The frequency $\mu  \sqrt{A_\ell}$ corresponds to the dispersion relation of a free conformally coupled scalar whose contribution vanishes in dimensional regularization, where one has
\be
\sum_{\ell =0}^\infty n_\ell \sqrt{A_\ell} = 0 \ .
\ee
Therefore, the main result of this section is 
\be
C_1  = r \delta E_1 + \frac{r}{2 \T} \sum_{\ell=0}^\infty \left(n_\ell \left[ (N-1)  \nu_{1,\ell} +  \nu_{2,\ell}\right] +2 q_{1,\ell} \nu_{1,\ell} + 2 q_{2,\ell} \nu_{2,\ell} \right)\ ,
\ee
where $\nu_{1,\ell}$ and $\nu_{2,\ell}$ are given respectively in \eqref{stab1} and \eqref{stab2}.

We now comment on the zero modes of the theory. It is shown that when a symmetry of the theory is broken by a classical field configuration, a given pair of stability angles vanishes \cite{dhn}. In our case, the stability angle $\nu_{2,0} $ corresponds to the zero mode related to the time translation symmetry on the cylinder stemming from the $t_0$ parameter in Eq.~\eqref{eq_phi_cyl}. The unnormalized eigenfunction is $\dot v(t)$ multiplied by the $\ell= 0$ spherical harmonic. Indeed, one has $\Lambda_2(\ell=0) = 1+4 m$, which coincides with one of the band edges of the $L_2$ Lam\'e operator $L_2$. To evaluate the associated stability angle, we use that for $\ell=0$ and $d=4$ the equations \eqref{eq:Translos} reduce to
\be
 \sn^2( \alpha_+ \,|\,  m ) = \frac{1}{m} \,, \qquad \sn^2(\alpha_- \,|\,  m)= 0 \ ,
\ee
and given the relations
\be
 -4 i  \KK(m) Z(\sn^{-1}(1/ \sqrt{m}  \,|\,  m) \,|\, m)=2 \pi \,, \qquad Z(\sn^{-1}(0  \,|\,  m) \,|\, m)= Z(0  \,|\,  m) = 0 \,, 
\ee
which can be derived using Eq.~\eqref{useful}, we arrive at the expected result
\be \label{zerozero}
 \nu_{2,0} =0 \ .
\ee
\noindent 
Additionally, there are $N-1$ zero modes $\nu_{1,0}$ associated with the coset $O(N)/O(N-1) \sim S^{N-1}$ corresponding to the symmetry breaking pattern induced by the classical solution. In fact, by evaluating Eq.~\eqref{lambda12} for $\ell=0$ and $d=4$ one has $\Lambda(\ell=1) = 1$, which lies at the band edge of the $\kappa=1$ Lam\'e operator and since
\be
\sn^{-1}\left(\left. 1\right|m\right) = \KK(m) \ ,
\ee
we have
\be
Z\left(\KK(m) | m\right) = 0 \ ,
\ee
implying
\be \label{zenzeroON}
\nu_{1,0} = 0 \ .
\ee

\subsection{Leading quantum correction: $C_1$} \label{ciuno}

We start by considering the sum over the stability angles of the $L_2$ operator, which yields the full contribution in the Ising model ($N=1$) case  
\be \label{finalform}
\frac{1}{2} \sum_{\ell=0}^\infty n_\ell \nu_{2,\ell}=   2 \T/r + \frac{1}{2} \sum_{\ell=2}^\infty n_\ell \nu_{2,\ell} \ ,
\ee
where we separated the contribution of the $\ell=1$ mode, which, as we shortly see, corresponds to a descendant state and is proportional to the period
\be
\nu_{2,1} = \T/r \ ,
\ee
as can be seen by evaluating Eq.~\eqref{stab2}  for $\ell = 1$. To compute the last term on the RHS, we subtract the UV divergent powers of $\ell$ in $d=4$ and add them back in the $\zeta$-regularized form. Therefore, we expand the summand as
\be
n_\ell \nu_{2,\ell} = \sum_{k=1}^\infty c_k l^{d-k} \ .
\ee
The first five terms diverge in $d=4$, and we subtract them. The resulting finite sum can be evaluated directly in $d=4$ dimensions, yielding
\be  \label{finalformm}
\frac{1}{2} \sum_{\ell=2}^\infty n_\ell \nu_{2,\ell}  =  \left[  \frac{1}{2}\sum_{\ell=2}^\infty \left(n_\ell \nu_{2,\ell} - \sum_{k=1}^5 c_k l^{d-k}\right)\rvert_{d=4} +\frac{1}{2} \sum_{k=1}^5 c_k \zeta(k-d)  -\frac12\sum_{k=1}^5 c_k\right] \ .
\ee 
The second term contains a divergent $1/\epsilon$ pole in Dim-Reg stemming from $\zeta(5-d) \sim \frac{1}{\epsilon} + \gamma_E + \order{\epsilon}$ and evaluates to 
\begin{align}
     \frac{1}{2} \sum_{k=1}^5 c_k \zeta(k-d)&=\pi-\frac{3 ((m-1) (3 m-2) \mathbb{K}(m)+(4 m-2) \ELE(m))}{(1-2 m)^{3/2} \epsilon } \nonumber\\
     &- \frac{2 (m-2) \mathbb{K}(m)+6 \ELE(m)}{\sqrt{1-2 m}} + \mathcal{O}(\epsilon) \ .
\end{align}

The last term in Eq.~\eqref{finalformm} is the contribution of the subtracted term evaluated for $\ell =1$. It appears because our sum over $\ell$ starts from $\ell =2$ while the defining sum of the zeta functions starts from $\ell =1$. It gives
\begin{align}
    -\frac{1}{2} \sum_{k=1}^{5}c_k|_{d=4} = \frac{(m (13-31 m)+2) \mathbb{K}(m)+4 \pi  (1-2 m)^{3/2}-(18-36 m) \ELE(m)}{(1-2 m)^{3/2}} + \mathcal{O}(\epsilon) \ .
\end{align}

Collecting all the results, we can express the sum over the stability angles in terms of a finite sum over $\ell$
\begin{align}\label{finalising}
    \frac{1}{2} \sum_{l=1} n_\ell \nu_{2,\ell} &= -\frac{3 ((m-1) (3 m-2) \mathbb{K}(m)+(4 m-2) \ELE(m))}{(1-2 m)^{3/2} \epsilon } + \frac{1}{2} \sum_{l=2}^{\infty} \sigma_2(\ell)\nonumber\\
    &-\frac{((29-5 m) m-14) \mathbb{K}(m)-5 \pi  (1-2 m)^{3/2}+(24-48 m) \ELE(m)}{(1-2 m)^{3/2}}  + \mathcal{O}(\epsilon)\ ,
\end{align}
where
\begin{align}\label{eq:sigma2l}
    \sigma_{2}(\ell) &= n_\ell \nu_{2,\ell}|_{d=4} + \frac{2}{\ell(1-2m)^{3/2}} \bigg[6 \left(\ell^2+\ell+1\right) (2 m-1) \ELE(m)+\pi  \ell (\ell+1)^2 (1-2 m)^{3/2}\nonumber\\
    &+ \bigg(6 + 4 \ell - 2 \ell^3 (3 + \ell) + (-15 + 2 \ell (1 + \ell) (-1 + 2 \ell) (5 + 2 \ell)) m\nonumber\\
    &+(9-4 \ell (\ell+1) (2 \ell (\ell+2)-1)) m^2\bigg)\mathbb{K}(m)\bigg] \ .
\end{align}

Following analogous steps for the stability angles of $\cO_1$, we arrive at
\begin{align} \label{finalform2}
\frac{1}{2} \sum_{\ell=0}^\infty n_\ell \nu_{1,\ell}&= -\frac{(m-1) (3 m-2) \KK(m)+(4 m-2) \ELE(m)}{3 (1-2 m)^{3/2} \epsilon} \nonumber \\ &+\pi + \frac{2 (m \KK(m)-\ELE(m))}{\sqrt{1-2 m}} + \frac{1}{2}\sum_{\ell=1} \sigma_1(\ell) +\order{\epsilon} \ ,
\end{align}
where
\begin{align}\label{eq:sigma1l}
    \sigma_1(\ell) &= n_\ell \nu_{1,\ell}|_{d=4} +\frac{2}{3\ell(1-2m)^{3/2}} \bigg[3 \pi  \ell (\ell+1)^2 (1-2 m)^{3/2}+2 (3 \ell (\ell+1)+1) (2 m-1) \ELE(m)\nonumber\\
    &- \bigg(6 (\ell+1) (\ell+2) \ell^2+(5-6 \ell (\ell+1) (4 \ell (\ell+2)+1)) m-2\nonumber\\
    &+3 (4 \ell (\ell+1) (2 \ell (\ell+2)+1)-1) m^2\bigg)\mathbb{K}(m)\bigg] \ .
\end{align}
By comparing the sum of Eq.~\eqref{finalising} and Eq.~\eqref{finalform2} (multiplied by $N-1$) to Eq.~\eqref{polepole}, one can see that the divergent $1/\epsilon$ pole cancels in the renormalized combination $\mathcal{S}_{\rm cl}-\frac{1}{2} \sum_{\ell=0}^\infty n_\ell \left(\nu_{2,\ell} + (N-1) \nu_{1,\ell}   \right)$. Dividing by the period $\T$, adding our results for $\delta E_1$ in Eq.~\eqref{deltaE1}, and finally taking the $\epsilon \to 0$ limit, we arrive at
\begin{align}
    C_1 &= \frac{1}{8 (1-2 m)^2 \mathbb{K}(m)}\Bigg[\mathbb{K}(m) (m (-7 m N+26 m+3 N-70)+28)\nonumber\\
    &+2 \pi  (1-2 m)^{3/2} (N+4)+4 (2 m-1) (N+11) \ELE(m)\Bigg]\nonumber\\
    &+\frac{1}{8 \sqrt{1-2m}\KK(m)} \left ( \sum_{\ell=2} \sigma_2(\ell) + (N-1)\sum_{\ell=1} \sigma_1(\ell) + 2\sum_{\ell=1} \left(q_{1,\ell} \nu_{1,\ell} + q_{2,\ell} \nu_{2,\ell} \right) \rvert_{d=4}  \right) \ .
\end{align}
This final expression for $C_1$ is our main technical result for the $O(N)$ CFT. The sums over $\ell$ can be evaluated numerically for any value of $m=m(\lambda n)$ or analytically in the small/large $\lambda n$ regime. Next, we will show that the small $\lambda n$ limit of $C_1$ reproduces the diagrammatic expansion for the scaling dimensions of the composite operators in the traceless symmetric Lorentz representations while providing explicit expressions for higher order terms.

\subsection{Perturbative semiclassics}\label{sec: perturbativeSemiclassics}
We will now provide a practical algorithm for the evaluation of scaling dimensions of the various operators in the traceless-symmetric Lorentz representation in the perturbative limit. While in this section we display results up to order $\order{\epsilon^2}$, higher orders can be straightforwardly obtained and are reported in App. \ref{8loop}. As we have seen in the classical computation, the small 't Hooft coupling $\lambda n$ expansion coincides with the small $m$ one (see Eq.~\eqref{mla}) and reproduces the standard perturbative loop expansion. 
By evaluating the small $m$ expansion of the stability angles as explained in App.~\ref{smallnu} we obtain
\begin{align}\label{lasterm}
    \frac{1}{2} \sum_{\ell=1}^\infty \sigma_1(\ell) &= \frac{11 \pi m^2}{64} + \mathcal{O}\left(m^3\right)= 11 \pi \left( \frac{\lambda n}{16 \pi^2}\right)^2 + \mathcal{O}\left(\left( \frac{\lambda n}{16 \pi^2}\right)^3\right) \,, \\
    \label{lastterm2}
     \frac{1}{2} \sum_{\ell=2}^\infty \sigma_2(\ell) &= \frac{51 \pi m^2}{64}  + \mathcal{O}\left(m^3\right)=51 \pi \left( \frac{\lambda n}{16 \pi^2}\right)^2 + \mathcal{O}\left(\left( \frac{\lambda n}{16 \pi^2}\right)^3\right) \ , 
\end{align}
where we used Eq.~\eqref{mla} to rewrite the expressions in terms of $\lambda n$. The remaining summands are
\begin{align}\label{descendant1}
          \frac{ r \nu_{1,\ell}}{\mathcal{T}}\bigg|_{d=4} &= \ell -\frac{2 (3 \ell+1)}{\ell+1} \left(\frac{\lambda n}{16 \pi^2}\right) \nonumber\\
    &- \left( \frac{20}{\ell+1}+\frac{8}{(\ell+1)^3}+\frac{2}{\ell+2}+\frac{2}{\ell}-51 \right) \left(\frac{\lambda n}{16 \pi^2}\right)^2 + \mathcal{O}\left(\left(\frac{\lambda n}{16 \pi^2}\right)^3\right) \ ,
\end{align}
\begin{align}
    \frac{r  \nu_{2,\ell}}{\mathcal{T}}\bigg|_{d=4} &= \ell - \frac{6 (\ell-1)}{\ell+1}\left(\frac{\lambda n}{16 \pi^2}\right)\nonumber\\
    &-\left(\frac{36}{\ell+1}+\frac{72}{(\ell+1)^3}+\frac{18}{\ell+2}+\frac{18}{\ell}-51\right)\left(\frac{\lambda n}{16 \pi^2}\right)^2 + \mathcal{O}\left(\left(\frac{\lambda n}{16 \pi^2}\right)^3\right) \ .
\end{align}
Collecting all the results and evaluating them at the fixed point \eqref{chiodofisso} we finally arrive at
\begin{align} \label{eq: excited Delta ON}
    &\Delta_{n,q_\ell} = n C_0(\lambda_* n) + C_1(\lambda_* n)  + \order{1/n} = n\left(1-\frac{\epsilon}{2}\right)+ \sum_{\ell=1}^\infty (q_{1,\ell} +q_{2,\ell}) \ell \nonumber\\
    &+ \bigg[\frac{3 n^2}{2 (N+8)} - \bigg(\frac{4-N}{2(N+8)} + \sum_{\ell=1}^{\infty} \frac{q_{1,\ell}(1+3\ell)+3 q_{2,\ell}(\ell-1)}{(1+\ell)(8+N)}\bigg)n + \mathcal{O}\left(n^0\right)\bigg] \epsilon\nonumber\\
    &+ \bigg[-\frac{17 n^3}{4 (N+8)^2} + \bigg(\frac{-11 N^2+10 N+604}{4 (N+8)^3}\nonumber\\
    &+\sum_{\ell=1}^{\infty} \frac{q_{1,\ell} (3 \ell (\ell+2) (\ell (\ell (17 \ell+43)+35)+5)-4)}{4 \ell (\ell+1)^3 (\ell+2) (N+8)^2}\nonumber\\
    &+ \sum_{\ell=1}^\infty \frac{(3 (\ell-1) (\ell (\ell (\ell (17 \ell+78)+135)+98)+12)) q_{2,\ell}}{4 \ell (\ell+1)^3 (\ell+2) (N+8)^2}\bigg)n^2 + \mathcal{O}(n)\bigg]\epsilon^2 + \mathcal{O}\left(\epsilon^3\right) \,.
\end{align}
In appendix \ref{8loop} we present additional explicit results up to the $8$-loop order.

Let us now consider some examples to learn how different operators emerge from the different integer values of $q_{1,\ell}$ and $q_{2,\ell}$. 

\subsubsection{Ising CFT}
We start from the Ising model corresponding to the $N=1$ case, where only integers $q_{2,\ell}$ occur. In this case Eq.~\eqref{eq: excited Delta ON} reduces to
\begin{align}\label{eq: ising general scaling dim}
&\Delta_{n,q_\ell }^{N=1} = n\left(1-\frac{\epsilon}{2}\right) + \sum_{\ell=1}^{\infty} q_{2,\ell} \ell + \frac{\epsilon}{6}   \left[n^2-2\left(\frac{1}{2}+\sum_{\ell=1}^{\infty} \frac{(\ell-1)q_{2,\ell}}{\ell+1}\right)n+ \order{n^0}\right]       \\ & - \frac{\epsilon^2}{324}  \Bigg[17 n^3- \left(67 +3 \sum_{\ell=1}^{\infty} \frac{ (\ell-1) \left(17 \ell^4+78 \ell^3+135 \ell^2+98 \ell+12\right) q_{2,\ell}}{\ell (\ell+1)^3 (\ell+2)}\right)n^2 + \order{n}\Bigg]  + \order{\epsilon^3} \ .   \nonumber   
\end{align}
This result appeared first in \cite{Antipin:2025ilv}.

By setting $q_{2,\ell}=0$ we obtain the dimension of the ground state operator, which we henceforth simply denote as $\Delta_n$
\begin{align}
    \Delta_n = n\left(1-\frac{\epsilon}{2}\right)+\frac{1}{6} (n-1) n \epsilon -\frac{n^2 (17 n-67)\epsilon ^2}{324}  + \mathcal{O}\left(\epsilon^2 n, \epsilon^3\right) \ ,  
\end{align}
which matches the diagrammatic result given in  Eq.~\eqref{checkitout} with $N=1$ for the tower of operators of the form $\phi^n$. In the first row of \autoref{table1}, we display the semiclassical result for these anomalous dimensions $\gamma_{n, q_\ell}$, defined as in Eq.~\eqref{anomalia}, together with the corresponding value of $q_{2,\ell}$. 

\renewcommand{\arraystretch}{1.5}

\begin{table}[] 
\centering
\begin{tabular}{|c|l|l|}
\hline
$q_{2,\ell}$                                         & Operators                                                                                   & Anomalous dimension $\gamma_{n, q_\ell}$                                                                                              \\ \hline
$0$                                                  & $\phi^n$                                                                                    & $\frac{n \epsilon  (3 n+N-4)}{2 (N+8)} +\frac{n^2 \epsilon ^2 (-17 n (N+8)+N (10-11 N)+604)}{4 (N+8)^3}$                                   \\
$\delta_{\ell,2} $                                   & $\partial^2 \phi^n $                                                                        & $ \frac{n \epsilon  (3 n+N-6)}{2 (N+8)}+\frac{n^2 \epsilon ^2 (-102 n (N+8)+N (197-66 N)+4720)}{24 (N+8)^3}$                               \\
$\delta_{\ell,3} $                                   & $\partial^3 \phi^n $                                                                        & $\frac{n \epsilon  (3 n+N-7)}{2 (N+8)}+\frac{n^2 \epsilon ^2 (-680 n (N+8)+N (1651-440 N)+34168)}{160 (N+8)^3}$                            \\
$\delta_{\ell,4} $                                   & $\partial^4 \phi^n$                                                                         & $\frac{n \epsilon  (15 n+5 N-38)}{10 (N+8)}+\frac{n^2 \epsilon ^2 (-4250 n (N+8)+N (11431-2750 N)+222448)}{1000 (N+8)^3}$                  \\
$\delta_{\ell,5} $                                   & $\partial^5 \phi^n $                                                                        & $\frac{n \epsilon  (3 n+N-8)}{2 (N+8)}+\frac{n^2 \epsilon ^2 (-1785 n (N+8)+N (5092-1155 N)+95756)}{420 (N+8)^3}$                          \\
$\delta_{\ell,6} $                                   & \textcolor{red}{$\partial^6 \phi^n$}                                    & $\frac{n \epsilon  (21 n+7 N-58)}{14 (N+8)}+\frac{n^2 \epsilon ^2 (-23324 n (N+8)+N (69145-15092 N)+1272088)}{5488 (N+8)^3}$               \\ \hline
\multirow{2}{*}{$2 \delta_{\ell,2} $}                & $\partial^4 \phi^n , \ \partial^2 \Box \phi^n ,$                                            & \multirow{2}{*}{$\frac{n \epsilon  (3 n+N-8)}{2 (N+8)}+\frac{n^2 \epsilon ^2 (-51 n (N+8)+N (167-33 N)+2908)}{12 (N+8)^3}$}                \\
                                                     & $\Box^2 \phi^n$                                                                           &                                                                                                                                            \\ \hline
\multirow{2}{*}{$\delta_{\ell,2} + \delta_{\ell,3}$} & $\partial^5 \phi^n ,  \partial^3 \Box \phi^n , $                                            & \multirow{2}{*}{$\frac{n \epsilon  (3 n+N-9)}{2 (N+8)}+\frac{n^2 \epsilon ^2 (-2040 n (N+8)+N (7693-1320 N)+124424)}{480 (N+8)^3}$}        \\
                                                     & $\partial \Box^2 \phi^n$                                                                  &                                                                                                                                            \\ \hline
\multirow{2}{*}{$\delta_{\ell,2}+\delta_{\ell,4} $}  & \textcolor{red}{$\partial^6 \phi^n$} ,  $\partial^4 \Box \phi^n ,$       & \multirow{2}{*}{$\frac{n \epsilon  (15 n+5 N-48)}{10 (N+8)}+\frac{n^2 \epsilon ^2 (-6375 n (N+8)+N (25709-4125 N)+402172)}{1500 (N+8)^3}$} \\
                                                     & $\partial^2 \Box^2 \phi^n$                                                                  &                                                                                                                                            \\ \hline
\multirow{2}{*}{$2 \delta_{\ell,3}$}                 & \textcolor{red}{$\partial^6 \phi^n$} ,  $\partial^4 \Box \phi^n , $      & \multirow{2}{*}{$\frac{n \epsilon  (3 n+N-10)}{2 (N+8)}+\frac{n^2 \epsilon ^2 (-340 n (N+8)+N (1451-220 N)+22088)}{80 (N+8)^3}$}           \\
                                                     & $\partial^2 \Box^2 \phi^n , \ \Box^3 \phi^n$                                                &                                                                                                                                            \\ \hline
\multirow{2}{*}{$ 3 \delta_{\ell, 2} $}              & \textcolor{red}{$\partial^6 \phi^n$} , $\partial^4 \Box \phi^n (2) ,  $ & \multirow{2}{*}{$ \frac{n \epsilon  (3 n+N-10)}{2 (N+8)}+\frac{n^2 \epsilon ^2 (-34 n (N+8)+N (157-22 N)+2304)}{8 (N+8)^3}$}               \\
                                                     & $\partial^2 \Box^2 \phi^n (3) , \ \Box^3 \phi^n$                                            &                                                                                                                                            \\ \hline
\end{tabular}
\caption{The table shows the $\order{\epsilon^2}$ anomalous dimensions for various operators of the Ising CFT along with the values of $q_\ell$ employed to obtain them via Eq.~\eqref{eq: ising general scaling dim}. We do not display terms that would be NNLO or higher-order in the semiclassical expansion and that are $\order{\epsilon n^0}$ at $1$-loop and $\order{\epsilon^2 n}$ at two loops. In the second column, the numbers between round brackets denote the total multiplicity of operators with the same given schematic form and the degenerate dimension to the semiclassical NLO. For $N=1$ these results are in agreement with the diagrammatic $1$-loop results reviewed in App.~\ref{add1loop}.} \label{table1}
\end{table}

To build the spectrum of excited states, we note that the stability angles are ordered as $\nu_{2,\ell+1} > \nu_{2,\ell}$ and stress that $q_{2,\ell}$ for each $\ell$ has a given integer value. Intuitively, exciting an $\ell =\ell^*$ mode adds $\ell^*$ derivatives, in the $(\ell^*/2,\ell^*/2)$ representation of $SO(3,1) \cong SL(2,\mathbb{C})/Z_2$, to the operator. In particular, as evident from \eqref{eq: ising general scaling dim}, for an operator of the form $\partial^s \Box^p \phi^n$ one has $\sum_{\ell=1}^\infty q_{2,\ell} \ell= 2p+s$ \footnote{For the $O(N)$ model this relation trivially generalizes as $\sum_{\ell=1}^\infty \left( q_{1,\ell} + q_{2,\ell} \right) \ell= 2p+s$.}. For example, if  a given  mode $\ell^\ast$ is excited twice and no other $\ell$ mode is excited we  have $q_{2,\ell} = 2 \delta_{\ell,\ell^\ast}$ with $\delta_{\ell,\ell^\ast}$ the Kronecker delta.  From the group theory point of view, we have to take the tensor product of the corresponding $(\ell^*/2,\ell^*/2)$ representations with itself and perform the decomposition into irreducible representations to determine the representations of the resulting operators. As a consequence, all the so-obtained representations remain degenerate to NLO in the semiclassical expansion.

\begin{itemize}
\item $q_{2,\ell}=\delta_{\ell,1}$ :
The first excited state is constructed by adding a single $\ell=1$ mode. Since $\nu_{2,1} = \T$ the corresponding scaling dimension increases by one unit, thereby yielding a descendant state. Therefore, a necessary (but not sufficient) condition for an excited state to be a primary is $q_{2,1} =0$. Moreover, given any primary operator with dimension $\Delta_{n,q_\ell}$, the corresponding full conformal multiplet of descendants is generated by exciting $\ell=1$ mode repeatedly. 

\item $q_{2,\ell}=\delta_{\ell,s}$ : In general, exciting a single $\ell=s$ mode yields the scaling dimension of operators of the form $\partial^s \phi^n$ 
\begin{align}
\label{spintower}
\Delta_{n,\delta_{\ell,s}}&=  n\left(1-\frac{\epsilon}{2}\right) + s + \frac{1}{6}   \left[n^2-2\left(\frac{1}{2}+\frac{s-1}{s+1}\right)n  \right] \epsilon - \frac{1}{324}  \Bigg[17 n^3 \nonumber \\ & - \left(67 + \frac{ 3(s-1) \left(17 s^4+78 s^3+135 s^2+98 s+12\right)}{s (s+1)^3 (s+2)}\right)n^2 \Bigg]\epsilon ^2  + \mathcal{O}\left(\epsilon n^0, \epsilon^2 n, \epsilon^3\right) \,.
\end{align}
This tower of spin $s$ operators corresponds to the single $\left(s/2,s/2 \right)$ representation and therefore is non-degenerate. First five members of this tower for $s=2,3,4,5,6$ are shown in the rows $2-6$ of \autoref{table1}. Recently in Ref.~\cite{Henriksson:2025vyi} our result Eq.~\eqref{spintower} for $s=2$ was combined with available perturbative data to derive the full two-loop conformal dimensions for the whole tower of spin-2 operators of the form $\partial^2 \phi^n$  for $n\geq 2$:
\begin{equation}
\Delta_{n, \delta_{\ell,s}} = 2+n\left(1-\frac{\epsilon}{2}\right)+\frac{(n-2)(3n+1)}{18}\epsilon-\frac{(n-2)(102n^2-335n-235)}{1944}\epsilon^2+\mathcal{O}\left(\epsilon^3\right) \,.
\end{equation}
Notice that for $n=2$ we have the conserved energy-momentum $T_{\mu \nu}$ whose conformal dimension is $\Delta_{T_{\mu \nu}}=d=4-\epsilon$.

\item $q_{2,\ell}=2 \delta_{\ell,2}$ : Exciting an $\ell=2$ mode twice we have to take the tensor product of the two spin-2 representations (with highest weights $1$) $1 \otimes 1 = 2\oplus 1 \oplus 0$. Correspondingly, the spin-4 $\partial^4 \phi^n$, spin-2 $\partial^2 \Box \phi^n$, and spin-0  $\Box^2 \phi^n$ components will be degenerate to this semiclassical order with conformal dimension:
\begin{align}\label{eq: spintwo1}
   \Delta_{n,2\delta_{\ell,2}}=4+n\left(1-\frac{\epsilon}{2}\right)+\frac{n(3n-7)}{18}\epsilon-\frac{51n^3-338n}{972}\epsilon^2+ \mathcal{O}\left(\epsilon n^0, \epsilon^2 n, \epsilon^3\right) \,,
\end{align}
as shown in the seventh row of \autoref{table1}.
This matches the corresponding terms in the full $1$-loop conformal dimension for these operators listed in \autoref{tableII}. Again, in \cite{Henriksson:2025vyi} our result was combined with available perturbative data to derive full two-loop conformal dimensions for the tower of spin-0 operators of the form $\Box^2 \phi^n$  for $n\geq 4$.

\item  $q_{2,\ell}=\delta_{\ell,2} + \delta_{\ell,3}$: Exciting one $\ell=3$ and one $\ell=2$ modes corresponds to the highest weights decomposing as $\frac{3}{2} \otimes 1 = \frac{5}{2} \oplus \frac{3}{2} \oplus \frac{1}{2}$ leading to spin-5, spin-3 and spin-1 components degeneracy with the result
\begin{align}
    \Delta_{n,\delta_{\ell,2} + \delta_{\ell,3}} = 5+n\left(1-\frac{\epsilon}{2}\right)+ \frac{n(3n-8) \epsilon}{18} -\frac{n^2 (2040 n-14533) \epsilon ^2}{38880} +\mathcal{O}\left(\epsilon n^0, \epsilon^2 n, \epsilon^3\right) \,,
\end{align}
shown in the eighth row of \autoref{table1} and matching the known $1$-loop results for the operators of the form $\partial^5 \phi^n, \ \partial^3 \Box \phi^n$, and $\partial \Box^2 \phi^n$ \cite{Kehrein:1994ff, Henriksson:2022rnm}, see \autoref{tableII}. 

It should be clear by now that exciting an $\ell^*$ mode corresponds to adding $\ell^*$ partial derivatives. Then, in the above example, the spin-one operator is built by fully contracting four out of five derivatives, obtaining a $\Box^2$ factor. Hence, one may identify the corresponding operator as $\del \Box^2 \phi^n$. Generically, given a specific set of $q_{2,\ell}$, there will be multiple operators with the same schematic form and degenerate dimension to the semiclassical NLO as we will see in our next example. 
\item $q_{2,\ell}=3 \delta_{\ell, 2}$ :
Exciting $\ell = 2$ mode three times corresponds to the highest
weights decomposing as $1 \otimes 1 \otimes 1 = 3\oplus  2 \oplus  1 \oplus 0$ with the representation $2$ occurring two times and $1$ occurring three times.  This yields the spin-$6,4,2,0$ operators shown in the last row of \autoref{table1} together with their corresponding multiplicity.

\end{itemize}

Evidently, by going further in our construction above one reproduces the full spectrum of primary operators in the traceless symmetric Lorentz representations whose spin is parametrically smaller than $n$. Remarkably, the framework naturally explains degeneracies in the Ising CFT spectrum \cite{Kehrein:1994ff, Kehrein:1995ia} which are harder to understand from a diagrammatic point of view. In particular, it is clear that the whole spectrum of operators becomes degenerate at the leading order in the semiclassical expansion, that is, in the $n \to \infty$ limit. In other words, perturbatively, the coefficient of the leading power of $n$ at any loop order is the same for all the families of operators and is equal to 1/6 and -17/324 at the one and two-loop orders, respectively (see Eq.~\eqref{eq: ising general scaling dim}). The NLO  semiclassical correction breaks the degeneracy among operators that can be distinguished by the choice of the $q_\ell$ integers.  Remaining  degeneracies are generically lifted by  higher orders in the semiclassical expansion; see \autoref{tableII} for the NNLO $\mathcal{O}\left(n^0\right)$ terms at $1$-loop order. 

Finally, we remark that our results correspond to the eigenvalues of the Hamiltonian on the cylinder (i.e., energy spectrum) which via the state-operator correspondence are mapped to the eigenvalues of the dilatation operator on the plane (spectrum of anomalous dimensions). From the standard perturbative diagrammatic expansion point of view,  before comparing to our results, we have to solve the operator mixing problem. For example, at spin-$6$ level and to $1$-loop in perturbation theory, for sufficiently large $n$, there are four towers of operators (shown in red color in  \autoref{table1}), three of which mix through a degree-3 polynomial, and one which remains rational. To compare with our results, the general expressions for the eigenvalues of the $3\times 3$ mixing matrix have to be expanded at large $n$, and the leading $\order{n^2}$ and subleading $\order{n}$ terms will be reproduced by our computation. The reason for this expansion is again the fact that the full $1$-loop expression for the conformal dimensions of the spinning excited states has the leading $\mathcal{O}\left(n^2\right)$, subleading $\mathcal{O}(n)$ \emph{and subsubleading power $\mathcal{O}\left(n^0\right)$} term while the ground state operators $\phi^n$ and scalars $\Box^2 \phi^n$ have only leading $n^2$ and subleading $n$ powers, see \autoref{tableII}.  As this sub-sub-leading $\mathcal{O}\left(n^0\right)$ term is NNLO in our semiclassical expansion, the full $1$-loop expression for the conformal dimension cannot be compared directly to our results\footnote{We thank J. Henriksson for  cross-checking the corresponding terms in perturbative $1$-loop results with our predictions in \cite{Antipin:2025ilv}.}. 

\renewcommand{\arraystretch}{1.5}



\subsubsection{$O(N)$ CFT}
Moving to the general $O(N)$ model, the ground state energy is obtained setting $q_{1,\ell}=q_{2,\ell}=0$ in Eq.~\eqref{eq: excited Delta ON}  yielding
\begin{align}
    \Delta_n = n\left(1-\frac{\epsilon}{2}\right) +\frac{ (3n+N-4) n \epsilon }{2 (N+8)}+\frac{n^2 \epsilon ^2 (-17 n (N+8)+N (10-11 N)+604)}{4 (N+8)^3} + \mathcal{O}\left(\epsilon^2 n, \epsilon^3\right) \,,
\end{align}
which corresponds to the singlet operator $(\phi_a \phi_a)^{n/2}$. In the remaining part of this section, we will simply write $\phi^n$ as a shorthand for the $O(N)$ singlet $(\phi_a \phi_a)^{n/2}$ where $n$ has to be even.  

The $P_k$ coefficients of the perturbative expansion of $\Delta_n$ (see Eq.~\eqref{roperto}) are polynomials of degree $k+1$ with no constant terms. Hence, according to the procedure explained in the last bullet point of Sec.~\ref{canovaccio}, using the state-of-the-art diagrammatic results \cite{Henriksson:2025hwi, Kompaniets:2017yct} we can determine, for the first time, the full three-loop contribution $P_3(n)$ to $\Delta_n$,  extending the current known order which is limited to $P_2(n)$  given in  Eq.~\eqref{checkitout}: 
\begin{align}
    \Delta_n^{\text{3-loop}} = n+ P_1(n)\epsilon+  P_2(n) \epsilon ^2 + P_3(n) \epsilon ^3 +\mathcal{O}\left(\epsilon^4\right) \ ,
\end{align}
\begin{align}   \label{3loopOn}
   &P_3(n)= \frac{n}{16 (N+8)^3} \Bigg[375  n^3 +  \frac{(32 (N+8) (N+26) \zeta_3   +3 N (95 N-312)-21648)n^2}{(N+8)}  \nonumber \\ & + \frac{2 n\left(9 N^4-766 N^3-8028 N^2  +11712 N+200032  +144 (N+8) ((N-7) N-102) \zeta_3\right)}{(N+8)^2}    \nonumber \\ & + \frac{3(-13 N^4+790 N^3+9104N^2+10784N-96256)-32 (N+8) (N (37 N+106)-872) \zeta_3}{(N+8)^2} \Bigg] \,. 
\end{align}

We  now excite $q_{1,\ell}$ modes while keeping $q_{2,\ell}=0$. Notice from  Eq.~\eqref{descendant1} that $\nu_{1,1} \neq \T$, so that exciting $\ell=1$ modes does not lead to the descendant states. This reflects the fact that for $N>1$ there are primary operators of the form $\Box \phi^n$ and $\partial \phi^n$ which are absent for $N=1$. Let us start by discussing the effect of these modes.
\begin{itemize}
    \item  $q_{1,\ell}= \delta_{\ell,1}$ : Exciting it once, we may, in principle, obtain singlet operators of spin-1 which are virial current candidates. The existence of a virial current,
i.e., a conserved spin-1 singlet operator with $\Delta=d-1$, is a necessary criterion for a theory that is scale  but not conformally invariant. The leading operator of this type has been shown in \cite{Henriksson:2022rnm} to be $\partial \Box \phi^6$ with dimension $\Delta=9+\mathcal{O}(\epsilon)$, which is well
above the required value for a virial current and requires exciting more than just one $\ell=1$ mode.

\item $q_{1,\ell}= 2\delta_{\ell,1}$ : Exciting it twice leads to
\begin{align}
   & \Delta_{n,2 \delta_{\ell,1}} = 2+n\left(1-\frac{\epsilon}{2}\right)+\frac{3n^2+(N-12)n  }{2 (N+8)}\epsilon-\frac{\epsilon ^2 n^2 \left(51 n N+408 n+33 N^2-254 N-3604\right)}{12 (N+8)^3}  \nonumber\\
   & +\mathcal{O}\left(\epsilon n^0, \epsilon^2 n, \epsilon^3\right) \,.
\end{align}
The related tensor decomposition informs us that the families of operators with this scaling dimension are $\partial^2 \phi^n$ and $\Box \phi^n$. Indeed, the above matches the known \cite{Henriksson:2022rnm} $1$-loop scaling dimension of the Lorentz scalar, $O(N)$ singlet operators $\Box \phi^n, \ n =4,6,8,...$ which we report in Eq.~\eqref{eq: known dev}. 
\item Another interesting limit  is $N\to \infty$ where we may connect to the description of operators in the large
$N$ expansion. For $n=4$, for example, there is the tower of spinning singlet operators of the form $[\sigma,\sigma]_\ell\sim \sigma \partial^\ell \sigma$, $\ell=2,4,6...\ $, with $\Delta_4^{[\sigma,\sigma]_\ell}=4+\ell+ \mathcal{O}(1/N)$ (see Eq. (6.13) of \cite{Henriksson:2022rnm} and Table 20 therein) where $\sigma \sim \phi_a \phi_a$ is the auxiliary Hubbard-Stratonovich field with $\Delta_\sigma=2+\mathcal{O}(1/N)$ introduced in the $1/N$ expansion of the model. Notice that the leading $N^0$ term in $\Delta_4^{[\sigma,\sigma]_\ell}$ is $\epsilon-$independent. To reproduce this operator tower we notice that if we neglect the missing NNLO terms, the leading $N^0$ prediction of Eq.~\eqref{eq: excited Delta ON} is also $\epsilon-$independent and reads
\begin{equation}
    \Delta_{n, q_\ell}  \overset{N \to \infty}{=} n+ \sum_{\ell=1}^\infty (q_{1,\ell} +q_{2,\ell}) \ell+ \mathcal{O}(1/N)  \,,
\end{equation}
which reproduces the  $\sigma \partial^\ell \sigma$ tower by fixing $\sum_{\ell=1}^\infty (q_{1,\ell} +q_{2,\ell}) \ell=\ell$ and $n=4$. Thereby, it looks reasonable to assume that for these operators the missing semiclassical orders scale at least as $1/N$. On the other hand, the neglected NNLO term at $1$-loop is expected to generate the leading $N^0$ dependence contributing to $\mathcal{O}(\epsilon)$ shift in conformal dimensions $4+\ell+ \mathcal{O}(\epsilon)+\mathcal{O}(1/N)$ required to reproduce the conformal dimensions of the other large-$N$ operators at spin-$\ell$.

\end{itemize}

\subsection{On the instabilities at large $\lambda n$} \label{Instambul}

\begin{figure}[t!]
\centering
\includegraphics[width=0.99\textwidth]{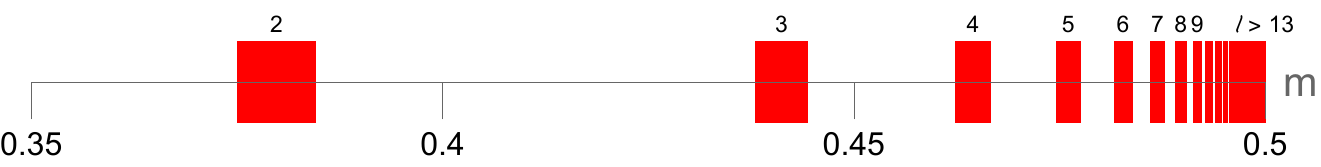} 
\caption{Values of $m$ for which complex stability angles occur. The numbers above the red intervals denote the corresponding value of $\ell$. For graphical reasons we do not display the numbers for the obvious $\ell=10, 11, 12$ modes.} 
	\label{complex}
\end{figure}

Building on the initial observation of \cite{Antipin:2025ilv}, we here discuss the instabilities of the classical orbit emerging at strong values of $\lambda n$. As discussed, a nonzero imaginary part in the stability angles signals an instability of the classical orbit. Starting from the $N=1$ case, we note that $\nu_{2,\ell}$ in Eq.~\eqref{stab2} becomes complex when $\ell$ falls in the range
\be \label{badrange}
 \sqrt{\frac{4-5 m}{1-2 m}}\le \ell + 1\le \sqrt{2}\sqrt{1+\frac{ \sqrt{1-m+m^2}}{1-2 m}} \ .
\ee
Physically, for these values of $\ell$, the $\Lambda_2(\ell)$ parameter takes values outside the allowed bands of the $L_2$ Lam\'e operator, as can be seen comparing Eq.~\eqref{lambda12} to Eq.~\eqref{2lamegap}. For small values of $\lambda n$, this condition cannot be satisfied by any integer $\ell$, and the instability does not manifest itself perturbatively. However, as $\lambda n$ increases, the $\ell=2$ stability angles become complex for  $3/8<m<5/13$ ($50 \lesssim \lambda n \lesssim 57$) where the naive semiclassical calculation breaks down. Nonetheless, for $5/13<m< \frac{1}{2}-\frac{1}{2 \sqrt{65}}$ all the stability angles are real again, with the next unstable modes being the $\ell=3$ one, which occurs for $\frac{1}{2}-\frac{1}{2 \sqrt{65}}<m<4/9$. The pattern repeats for a while as can be seen from Fig. \ref{complex}, showing the values of $m$ for which there are complex stability angles which can be obtained by rewriting Eq.~\eqref{badrange} as a condition on $m$.
Finally, there will always be at least one unstable mode for any $m>\frac{1}{2}-\frac{1}{2 \sqrt{12545}}$ corresponding to $\lambda n \approx 9149$. 
By recalling that for $\lambda n \to \infty$ we have $m  \to 1/2$, one can see that the set of stability angles tends to a continuum with the complex ones filling the whole region $\frac{\sqrt{3/2}}{\sqrt{1-2m }}<  \ell < \frac{3^{1/4}}{\sqrt{1-2m }}$. Consequently, one can make the instability manifest by investigating the large $\lambda n$ limit of $C_1$. For the sake of simplicity, we focus on the ground state operator and set $q_{2,\ell}=0$. To determine the large $\lambda n$ expansion of $C_1$ we introduce
\be
M =\frac{1}{\sqrt{1-2 m}} \in [1, \infty) \ ,
\ee
mapping the large $\lambda n$ limit into large $M$. In particular, using Eq.~\eqref{largelambdanm}, we have
\be \label{largeM}
M \sim \left(\frac{3 \Gamma \left(\frac{3}{4}\right)}{\sqrt{2} \pi ^{3/2} \Gamma \left(\frac{1}{4}\right)} \right)^{1/3} (\lambda n)^{1/3} +\order{ (\lambda n)^{-1/3}} \ .
\ee
To evaluate the sum over $\ell$ in Eq.~\eqref{finalising} in the large $M$ limit, we follow \cite{Cuomo:2021cnb, Antipin:2022naw} and split it as\footnote{For technical convenience, we include the $\ell=1$ mode in the sum.} 
\be \label{bananasplit}
\frac12\sum_{\ell=1}^\infty \sigma_2 (\ell) = \frac12\sum_{\ell=1}^{\Lambda M} \sigma_2 (\ell) +\frac12\sum_{\Lambda M+1}^\infty \sigma_2 (\ell) \,,
\ee
where $\Lambda$ is an arbitrary cutoff chosen such that $\Lambda M$ is an integer.  For the first sum in the RHS of Eq.~\eqref{bananasplit}, we obtain 
\be \label{lowmodes}
\frac12\sum_{\ell=1}^{\Lambda M} \sigma_2 (\ell) = \frac{3 M^4}{16} (\log (\Lambda M)+\gamma_E) +\order{M^3} \,.
\ee
To evaluate the second sum in the RHS of Eq.~\eqref{bananasplit}, we introduce $k \equiv \ell/M$ and consider the Euler-Maclaurin summation formula
\be \label{eul}
 \sum_{\Lambda M+1}^\infty \sigma_2 (\ell)  =M\int_{\Lambda}^\infty dk \ \Sigma(k)-{\frac{\Sigma(\Lambda)}{2}}-\sum_{i=1}{\frac{B_{2i}}{(2i)! M^{2i-1}}}\, \Sigma^{(2i-1)}(\Lambda)\,, \quad
\Sigma(k)\equiv  \sigma_2 (k M)\,,
\ee
where $B_{2i}$ denotes the $2i$-th Bernoulli number. Since the remainder terms are suppressed by increasing powers of $M$, at the leading order, only the integral contributes. This reflects the fact that the stability angles form a continuum in this limit. Therefore, by expanding $\Sigma(k)$ as $\Sigma(k) = M^3 \tilde{\Sigma}(k) + \order{M^2}$, we arrive at
\be
 \sum_{\Lambda M+1}^\infty \sigma_2 (\ell)  =M^4\int_{\Lambda}^\infty dk \ \tilde{\Sigma}(k) + \order{M^3} \,,
\ee
where 
\begin{align}
\tilde{\Sigma}(k) &= \frac{\sqrt{2 \pi } \Gamma \left(\frac{3}{4}\right) k^2}{\Gamma \left(\frac{1}{4}\right)}+\frac{-8 k^4+12 k^2+3}{16 k}-\frac{3  \Gamma \left(\frac{3}{4}\right) \ELE(-1) k}{\sqrt{\pi } \Gamma \left(\frac{1}{4}\right)}+\frac{k^2}{2 \sqrt{6}}\Bigg[\frac{1}{\Gamma \left(\frac{1}{4}\right)} \sqrt{\frac{\sqrt{9-3 k^4}+k^2}{\sqrt{9-3 k^4}+k^2-3}} \nonumber  \\  & \times\sqrt{\sqrt{9-3 k^4}+k^2+3} \left(\Gamma \left(\frac{1}{4}\right)-2 \sqrt{\frac{2}{\pi }} \Gamma \left(\frac{3}{4}\right) \Pi \left(\frac{1}{6} \left(-k^2-\sqrt{9-3 k^4}+3\right)|\frac{1}{2}\right)\right)\nonumber \displaybreak \\ & +\frac{\sqrt{\left(2 k^2+3\right) \left(\sqrt{9-3 k^4}+k^2-3\right)} \left(\Gamma \left(\frac{1}{4}\right)-2 \sqrt{\frac{2}{\pi }} \Gamma \left(\frac{3}{4}\right) \Pi \left(\frac{1}{6} \left(-k^2+\sqrt{9-3 k^4}+3\right)|\frac{1}{2}\right)\right)}{\Gamma \left(\frac{1}{4} \right)\sqrt{\sqrt{9-3 k^4}-k^2+3} }\Bigg] \ .
\end{align}

We can isolate the cutoff dependence by subtracting the divergent infrared behavior as
\begin{align}
M^4\int_\Lambda^\infty dk  \tilde{\Sigma}(k) &= M^4\left(\int_0^\infty dk  \left(\tilde{\Sigma}(k) -\frac{3}{16 k (1 + k^2)} \right)+\int_{\Lambda}^\infty \frac{3}{16 k (1 + k^2)}dk \right) \nonumber \\ &=\int_0^\infty dk \left(\tilde{\Sigma}(k) -\frac{3}{16 k (1 + k^2)} \right)+ \frac{3}{32} \left(\log \left(\Lambda ^2+1\right)-2 \log \Lambda \right)    \,.
\end{align}
By combining the above with Eq.~\eqref{lowmodes}, it becomes possible to take the $\Lambda \to 0$ limit to obtain
\be
\frac12\sum_{\ell=1}^{\infty} \sigma_2 (\ell) = M^4 \left[\frac{3}{16} (\log M+\gamma_E) +\int_{0}^\infty dk  \left(\tilde{\Sigma}(k) -\frac{3}{16 k (1 + k^2)} \right)\right]  +\order{M^3} \,.
\ee
By substituting the above into Eq.~\eqref{finalising} and removing the contribution of the $\ell=1$ mode to avoid double-counting it, we arrive at
\be
C_1(n, q_{\ell}=0 )= M^4 \left[\frac{3}{16} \left(\log M+\gamma_E-\frac32 \right) +\int_{0}^\infty dk  \left(\tilde{\Sigma}(k) -\frac{3}{16 k (1 + k^2)} \right)\right]  +\order{M^3} \,.
\ee
Finally, by using Eq.~\eqref{largeM}, the fixed point value Eq.~\eqref{chiodofisso}, and adding the leading order contribution $C_0$ in Eq.~\eqref{Largeorder}, we obtain
\begin{align}
\Delta_n = \frac{1}{\epsilon} \left[\left(\frac{ \sqrt{\frac{3 \pi }{2}} \Gamma \left(\frac{3}{4}\right)}{\Gamma \left(\frac{1}{4}\right)}n \epsilon\right)^{\frac{4-\epsilon }{3-\epsilon }}\left(1+ \alpha \epsilon  + \order{\epsilon^2}\right) + \order{n^{\frac{2-\epsilon }{3-\epsilon }}} \right] \,,
\end{align}
where
\be
\alpha= \frac{1}{18} \left(6 \gamma_E -9+3\log \frac{4}{3}+ 32\int_{0}^\infty dk  \left(\tilde{\Sigma}(k) -\frac{3}{16 k (1 + k^2)} \right)\right) \,.
\ee
Interestingly, the NLO correction modifies the large $n$ behavior from $\Delta_n \sim n^{4/3}$ to $\Delta_n \sim n^{\frac{d}{d-1}}$ mimicking the generic non-perturbative behavior of the scaling dimension of the lowest operator with charge $n$ in generic CFTs with global symmetries \cite{Hellerman:2015nra}. As should be clear from the previous discussion, in our case this is not the end of the story. In fact, $\tilde{\Sigma}(k)$ exhibits a branch cut in the region $\sqrt{3/2}< k < 3^{1/4}$ corresponding to the large $\lambda n$  ($m \to 1/2$) limit of Eq.~\eqref{badrange}. The latter gives rise to an imaginary contribution to the energy which is related to the decay rate of the state on the cylinder. These instabilities and their consequence on the CFT spectrum deserve a delicate analysis that will be pursued in future work.

Moving to the general $O(N)$ CFT, we conclude this section by noting that the $\nu_{1.\ell}$ modes also become complex when  
\be \
\ell \le \frac{1-2 m-\sqrt{2 m^2-3 m+1}}{2 m-1} \,,
\ee
implying that the instability survives in the large $N$ limit where these modes provide the leading contribution due to their multiplicity $N-1 \sim N$. As $\lambda n$ increases, the first complex stability angle appears at $m =3/7$ corresponding to $\lambda n \approx 128$.

\section{The $O(N)$ $\phi^6$ theory in $d=3-\epsilon$} \label{phi6}

Let us now turn our attention to our second example, namely the $O(N)$ $\phi^6$ model in $d=3-\epsilon$ dimensions with Lagrangian
\begin{equation}
  \mathcal{L} = \frac{1}{2}\left( \partial \phi_a \right)^2 - \frac{\lambda^2}{6} (\phi_a\phi_a)^3  \,, \quad a=1,\dots,N \,,
\end{equation}
which is known to develop a fixed point for the following  value of the coupling
\begin{equation}\label{fpphi6}
    \frac{\lambda_*^2}{4 \pi^2} = \frac{\epsilon}{3N+22} + \mathcal{O}\left(\epsilon^2\right) \ .
\end{equation}
A peculiar feature of this model is that the beta function of the sextic coupling vanishes at the one-loop level and, therefore, to this order, the theory is conformal in exactly three dimensions. As a consequence, there will be no $1/\epsilon$ pole (associated with a genuine UV divergence) arising when regularizing the sum over the stability angles in dimensional regularization. Additionally, the $\delta E_1$ contribution in Eq.~\eqref{finafinal} vanishes. Compared to the $\phi^4$ model, much less is known about the composite operator spectrum of the theory. In particular, we will be able to compare our results to perturbation theory only for the scaling dimensions of the $\phi^n$ operators, which for generic $n$, is known up to the four-loop ${\order\epsilon^2}$ order \cite{Bednyakov:2025usv, Hager:2002uq} and reads, 

\begin{align}\label{Deltaphi6}
    \Delta_n &= \frac{n}{2}(1-\epsilon)+\frac{n(n-2)   (5 n+3
   N-8) \epsilon}{6 (3 N+22)}+\frac{\epsilon^2}{192 (3 N+22)^3} \bigg(-1048 n^5 \Big(3 N+22\Big)\nonumber\\
   &+n^4 \Big(-16 \Big(71 N-576\Big) \Big(3 N+22\Big)-27 \pi ^2 \Big(N+24\Big) \Big(3 N+22\Big)\Big)\nonumber\\
   &+n^3 \Big(54 \pi ^2 \Big(N+24\Big) \Big(3 N+22\Big)+8 \Big(3 N \Big(\Big(802-35 N\Big) N+4796\Big)-66664\Big)\nonumber\\
   &-2 \pi ^2 \Big(N \Big(N \Big(22 N+595\Big)-3586\Big)-47656\Big)\Big)+n^2 \Big(8 \Big(N \Big(669 N^2-996 N-36020\Big)\nonumber\\
   &+64272\Big)-\pi ^2 N \Big(N \Big(N \Big(3 N-140\Big)-7448\Big)-26536\Big)+4 \pi ^2 \Big(N \Big(N \Big(22 N+595\Big)\nonumber\\
   &-3586\Big)-47656\Big)-118496 \pi ^2\Big)+n \Big(2 \pi ^2 N \Big(N \Big(N \Big(3 N-140\Big)-7448\Big)-26536\Big)\nonumber\\
   &-16 \Big(N \Big(N \Big(435 N+1792\Big)-10308\Big)+7856\Big)+236992 \pi ^2\Big)\bigg)+\cO(\epsilon^3)
\end{align}
Additionally, the coefficients of the leading and subleading powers of $n$ at the $6$-loop order have been recently calculated in \cite{Bednyakov:2025usv}, providing a nontrivial additional test of the semiclassical calculation.

\subsection{Classical solution and leading order: $C_0$}

Proceeding as explained in the previous sections, we perform a Weyl transformation to the cylindrical geometry, where, again assuming a ground state of the form $\phi_{1,\text{cl}} = v(t)$, $\phi_{i,\text{cl}} = 0 \,, \ i=2, \dots N $, the EOM reads
\begin{equation}\label{eq:carmelo}
\frac{d ^2v}{d t^2}+\mu^2 v +\lambda ^2 v^5= 0 \ ,
\end{equation}
with solution\footnote{Although this solution was derived independently, a related approach is discussed in \cite{Jia:2006xe}, where the authors describe how to obtain a $\phi^6$ field configuration from a $\phi^4$ field solution. Their procedure might also be applicable in the present context.} 
\begin{equation}\label{phi6fullsol}
    v(t) = \frac{a \,\text{cn}\left(\left.\omega t\right|m\right)}{\sqrt{1 + b \,\text{cn}\left(\left.\omega t\right|m\right)^2}} \,, \qquad \omega =\mu \left(1  +  m (m-1) \right)^{-1/4} \ ,
\end{equation}
where the coefficients $a$ and $b$ depend on $m$ as follows
\begin{equation}
    \begin{split}
            a &= (2\mu)^{\frac{1}{2}} \left(\frac{2 (m-1) m-\frac{(m-2) (m+1) (2 m-1)}{\sqrt{(m-1) m+1}}+2}{36 g^2 (m-1)^2} \right)^{1/4}  \ ,
    \end{split}
\end{equation}
and
\begin{align}
b &= \frac{2 m+\sqrt{(m-1) m+1}-1}{3 (1-m)} \ .
    \end{align}
In order to write a compact and manageable expression, we  analytically continued the theory to complex values of the sextic coupling by introducing $g=-i\lambda$, such that our solution is a real function for $m\in[0,1]$\footnote{It is also possible to work with complex $m$, but it is convenient to work with $m\in[0,1]$, as it allows us to use all the Jacobi functions relations and properties we need throughout the computation.}. From the coupling dependence of the solution evaluated at the fixed point \eqref{fpphi6}, we see that the semiclassical expansion assumes the form \eqref{dslOK} with $\beta=1/4$. The period of the solution is
 \be
\T = 4 \KK(m) / \omega  \,.
 \ee
To solve for the integral defining the action variable $I$ \eqref{eq:actionvar} entering the Bohr-Sommerfeld condition \eqref{BS}, it is helpful to introduce yet another quantity $u = \cn\left(\left.\omega t\right|m\right)$, to obtain
\begin{align}
I &= 16 \pi r^2 \omega a^2 \int \frac{\sqrt{1-u^2} \sqrt{m \left(u^2-1\right)+1}}{\left(b u^2+1\right)^3} \, du  \nonumber \\ &= -\frac{16 \pi r^2\omega a^2 }{8 b^2 (b+1) \sqrt{1-m} (b (m-1)+m)} \Bigg[ \left((3 b+4) b^3 \right.  \nonumber \\ & \left. -6 (b+1)^2 b^2 m+(b+1)^3 (3 b-1) m^2\right) \Pi \left(-b\left|\frac{m}{m-1}\right.\right) \nonumber \\ &-(b (3 b+2) (m-1)-m) (b (m-1)+m) \KK\left(\frac{m}{m-1}\right)   \nonumber \\ &+b (m-1) (b (3 b (m-1)+4 m-2)+m) \ELE\left(\frac{m}{m-1}\right)\Bigg] \,.
\end{align}
The solution of the Bohr-Sommerfeld condition \eqref{eq:actionvar} in the small $g n$ regime yields
\be\label{BSphi6}
m = \frac{4 }{\sqrt{3} \pi }(g n) -\frac{8 }{3 \pi ^2}(g n)^2 + \frac{5 \sqrt{3} }{\pi ^3} (g n)^3-\frac{148}{9 \pi^4} (gn)^4+\order{(gn)^6} \,.
\ee
By using this result, one can check that the small $\lambda n$ expansion of Eq.~\eqref{phi6fullsol} agrees with the perturbative solution found in \cite{Antipin:2024ekk}.
We can then compute the energy of the solution \eqref{phi6fullsol}, whose full expression is
\be
C_0 = \Omega_{d-1}r^d\braket{T_{00}} \rvert_{d=3} = \frac{\sqrt{2} \pi  (r\omega)^2}{3 g n}  \sqrt{1-m+m^2-2 (m-2) (m+1) (2 m-1) (r\omega) ^2} \ .
\ee
Using the solution \eqref{BSphi6} of the Bohr-Sommerfeld constraint, reverting back to $\lambda=ig$, and expanding for $\lambda n\ll1$, we get
\begin{equation}
    C_0=\frac{1}{2}+\frac{5}{24\pi^2}(\lambda n)^2-\frac{131}{384\pi^4}(\lambda n)^4+\frac{4915}{4608\pi^6}(\lambda n)^6+\order{(\lambda n)^8} \ ,
\end{equation}
which, at the fixed point \eqref{fpphi6} yields
\begin{equation}\label{eq:LOPhi6}
\begin{split}
    nC_0&=\frac{n}{2}+\frac{5  }{6 (3
   N+22)}n^3 \epsilon-\frac{131 }{24 (3 N+22)^2}n^5 \epsilon ^2+\frac{4915 }{72 (3 N+22)^3}n^7 \epsilon ^3 +\order{n^9 \epsilon^4} \ .
   \end{split}
\end{equation}
Finally, we give the large $\lambda n$ expansion of the classical contribution to $\Delta_n$
\begin{align}
n C_0 &= \frac{(-1)^{15/8} \pi  \sqrt{\lambda } }{ 3^{11/8} \sqrt{2} \KK\left(\frac{1-i \sqrt{3}}{2}\right)^{3/2}} n^{3/2} + \frac{(-1)^{31/24} \pi  }{2^{3/2} 3^{1/8} \KK\left(\frac{1-i \sqrt{3}}{2}\right)^{3/2} \sqrt{\lambda }}  \nonumber \\ &\times\Bigg(\KK\left(\frac{1-i \sqrt{3}}{2}\right)  -\Pi \left(\frac{3-i \sqrt{3}}{6} \Bigg|\frac{1-i \sqrt{3}}{2} \right)\Bigg)n^{1/2}+ \order{n^{-1/2}} \ .
\end{align}
in agreement with the generic expectation $\Delta_{n, q_\ell}\underset{n \to \infty}\sim n^{\frac{d}{d-1}}$. Despite appearances, the  expression above is real for all the real values of $\lambda$ and $n$.

\subsection{Leading quantum correction: $C_1$}
We now proceed with the computation of the quantum corrections to the leading semiclassical result. As explained in the previous sections, we do so by considering the fluctuations $\eta(\vec{x},t)$ and $\tilde{\phi}_a(\vec{x},t)$ of the classical solution as $\phi_1(\vec{x},t)=v(t)+\eta(\vec{x},t)$ and $\phi_a(\vec{x},t)=\tilde{\phi}_a(\vec{x},t) \,, \ a=2, \dots, N$ to obtain the quadratic Lagrangian
\begin{equation}
    \mathcal{L}_2=\frac{1}{2}\eta\mathcal{O}_{||}\eta + \frac{1}{2}\tilde{\phi}_a\mathcal{O}_{\perp}\tilde{\phi}_a\ ,
\end{equation}
where a sum is intended over the $a=2,\dots,N$ index.\\
Expanding the fluctuations $\eta$  and $\tilde{\phi}$ in a basis of spherical harmonics, we find for each component the quadratic operators
\begin{equation} \label{eq:quad_op_6}
    \mathcal{O}_{||,\ell}=-\partial^2_t-J^2_{\ell}-\mu^2-5\lambda^2v^4(t) \ ,
\end{equation}
\begin{equation} \label{eq:quad_op_6_2}
    \mathcal{O}_{\perp,\ell}=-\partial^2_t-J^2_{\ell}-\mu^2-\lambda^2v^4(t) \ .
\end{equation}
Unlike the $\phi^4$ case, the fluctuation operators appear to be rather cumbersome and we leave to future investigations the problem of finding the exact analytic expression of their stability angles. Instead, we compute them perturbatively in the small $\lambda n$ limit as outlined in Appendix \ref{smallnu6}. As explained in Sec.~\ref{canovaccio}, armed with their expression, we can compute the one-loop contribution to the scaling dimension according to the formula
\begin{equation}\label{eq:phi6excited}
    C_{1}=\frac{r}{2\T}\sum^{\infty}_{\ell=1}\left(n_{\ell}\Big[\nu_{||,\ell}+(N-1)\nu_{\perp,\ell} \Big]+2q_{||,\ell}\nu_{||,\ell}+2q_{\perp,\ell}\nu_{\perp,\ell}\right)\,,
\end{equation}
where $\nu_{||,\ell}$ are the stability angles relative to the operator $ \mathcal{O}_{||,\ell}$ and analogously $\nu_{\perp,\ell}$ are those relative to $ \mathcal{O}_{\perp,\ell}$. As for  the $\phi^4$ model, one has $\nu_{||,\ell=0}=0$, due to the zero mode associated with time translation symmetry on the cylinder, and $\nu_{||,\ell=1}=\T$ corresponding to the descendant states. \\
Order by order in the small $\lambda n$ expansion one can regularize the sum over $\ell$ using dimensional regularization following the same steps as for the $\phi^4$ model in Sec.~\ref{ciuno}. For the NLO correction to the ground state we obtain
\begin{equation}
\begin{split}
     C_1(n, q_{\ell}=0)\ &=\left(\frac{N-6 }{8 \pi ^2}\right)(\lambda n)^2-\left(\frac{1136 + 27 \pi^2}{3072 \pi ^4}N+\frac{27}{128 \pi ^2}-\frac{3}{\pi ^4}\right)(\lambda n)^4+\order{ \lambda^6 n^6} \ ,
\end{split}
\end{equation}
which at the fixed point \eqref{fpphi6} reads
\begin{equation}
\begin{split}
     C_1(n, q_{\ell}=0)&=\left(\frac{ N-6  }{6 N+44}\right)\epsilon n^2+\frac{ 9216-1136 N-27 \pi ^2 (N+24) }{192 (3 N+22)^2}\epsilon^2 n^4+\order{\epsilon^3 n^6} \ ,
\end{split}
\end{equation}
in agreement with Eq.~\eqref{Deltaphi6}. Taking into account the small $\lambda n$ expansion of the stability angles, which for $\ell > 1$ reads
\begin{equation}
    \begin{split}
          \frac{r \nu_{||,\ell}}{\T}& =\ell-\frac{5 (\ell-1) 
   }{(2 \ell+1) (3 N+22)}n^2 \epsilon\\
   &+\frac{5 (\ell (\ell+1) (\ell (2 \ell (4 \ell (\ell (262 \ell+325)-267)-1447)-647)+936)+90)}{12 \ell (2 \ell+1)^3 \left(4 (\ell+2) \ell^2+\ell-3\right) (3 N+22)^2} n^4 \epsilon ^2+\order{\epsilon^3 n^6} \ ,
    \end{split}
\end{equation}
\begin{equation}
    \begin{split}
          \frac{r \nu_{\perp,\ell}}{\T}& =\ell-\frac{(5 \ell+1)  }{(2
   \ell+1) (3 N+22)}n^2 \epsilon\\
   &+\frac{(5 \ell (\ell+1) (\ell (2 \ell (4 \ell (\ell (262 \ell+589)+261)-415)-695)-72)+18)}{12 \ell (2 \ell+1)^3 \left(4 (\ell+2) \ell^2+\ell-3\right) (3 N+22)^2} n^4 \epsilon ^2+\order{\epsilon^3 n^6} \ ,
    \end{split}
\end{equation}
we finally arrive at 
\begin{equation}\label{eq:Phi6final}
\begin{split}
& \Delta_{n,q_{\ell}}=n C_0(\lambda_*n)+C_1(\lambda_*n)+\order{1/n}=\frac{n}{2}+\sum_{\ell=1}^{\infty}(q_{||,\ell}+q_{\perp,\ell})\ell\\
    &+\Bigg[\frac{5  }{6 (3
   N+22)}n^3-\Bigg(\sum_{\ell=1}^{\infty}q_{||,\ell}\frac{5 (\ell-1) 
   }{(2 \ell+1) (3 N+22)}+\sum_{\ell=1}^{\infty}q_{\perp,\ell}\frac{5 \ell+1  }{(2
   \ell+1) (3 N+22)}\\
   &-\frac{ N-6  }{6 N+44}\Bigg)n^2+\cO(n) \Bigg]\epsilon-\Bigg[\frac{131 }{24 (3 N+22)^2}n^5-\Bigg(\frac{ -1136 N-27 \pi ^2 (N+24)+9216 }{192 (3 N+22)^2}\\
   &+\sum_{\ell=1}^{\infty}q_{||,\ell}\frac{450 + 4680 \ell + 1445 \ell^2 - 17705 \ell^3 - 
 25150 \ell^4 + 2320 \ell^5 + 23480 \ell^6 + 
 10480 \ell^7}{12 \ell (2 \ell+1)^3 \left(4 (\ell+2) \ell^2+\ell-3\right) (3 N+22)^2}\\
&+\sum_{\ell=1}^{\infty}q_{\perp,\ell}\frac{18 - 360 \ell - 3835 \ell^2 - 7625 \ell^3 + 
 6290 \ell^4 + 34000 \ell^5 + 34040 \ell^6 + 
 10480 \ell^7}{12 \ell (2 \ell+1)^3 \left(4 (\ell+2) \ell^2+\ell-3\right) (3 N+22)^2}\Bigg)n^4+\cO(n^3)\Bigg]\epsilon ^2\\
&+\order{\epsilon^3} \ .
\end{split}
\end{equation}
An explicit expression up to the eight-loop order  is presented in App.~\ref{8loop}, which, when restricted to six loops is in perfect agreement with  the result given in  \cite{Bednyakov:2025usv}.
As an additional check, we  test our result for the excited states in the case of the tricritical Ising theory (i.e., $N=1$) against the known results for the anomalous dimension at order $\order{\epsilon}$ for operators with spin. In fact, setting $q_{\perp,\ell}=0$ and $N=1$ in Eq.~\eqref{eq:Phi6final} we recover the leading and subleading terms of the $1/n$ expansion of the following $1$-loop anomalous dimensions computed in \cite{Henriksson:2025kws}:
\begin{itemize}
\item Operators of the form $\partial^2\phi^n$ with
\begin{equation}
\gamma_{n,\delta_{\ell,2}}  = \frac{(n-3)(n-2)(5n+4)}{150} \epsilon + \mathcal{O}\left(\epsilon^2\right)\,, \qquad n \ge 2 \,,
\end{equation}
are reproduced in \eqref{eq:Phi6final} by taking $q_{||,\ell}=\delta_{\ell,2}$;
\item Operators of the form $\partial^3\phi^n$ with
\begin{equation}
\gamma_{n,\delta_{\ell,3}} = \frac{(n-3)\left(7n^{2} - 12n - 40\right)}{210} \epsilon + \mathcal{O}\left(\epsilon^2\right)\,,
\qquad n \ge  3 \,,
\end{equation}
are reproduced in \eqref{eq:Phi6final} by taking $q_{||,\ell}=\delta_{\ell,3}$;
\item Operators of the form $\Box^2\phi^n$ with
\begin{equation}
\gamma_{n,2\delta_{\ell,2}} = \frac{(n-4)(n-2)(5n+3)}{150}\epsilon + \mathcal{O}\left(\epsilon^2\right) \,,
\qquad n \ge  4 \,,
\end{equation}
are reproduced in \eqref{eq:Phi6final} by taking $q_{||,\ell}=2\delta_{\ell,2}$;
\item Operators of the form $\partial^4\phi^n$ with
{\small
\begin{equation}
\gamma_{(\pm)} =
\frac{
35n^{3} - 182n^{2} + 32n + 684
\pm
\sqrt{49n^{4} + 812n^{3} + 37228n^{2} - 287184n + 508176}}{1050}\epsilon + \mathcal{O}\left(\epsilon^2\right) \,,
\end{equation}}
\noindent with $\gamma_{(+)}$ existing for $n\ge 2$ and $\gamma_{(-)}$ for $n \ge4$. Their expansion around $n\to\infty$ is reproduced, to   NLO, by taking $q_{||,\ell}=\delta_{\ell,4}$ for $\gamma_{(+)}$ and $q_{||,\ell}=2\delta_{\ell,2}$ for $\gamma_{(-)}$. These operators provide an interesting example as their $1$-loop anomalous dimension is not a polynomial in $n$;
\item Operators of the form $\Box^2\partial\phi^n$ with
\begin{equation}
\gamma_{n,\delta_{\ell,2}+\delta_{\ell,3}} =\frac{35n^3-207n^2+64n+600}{1050}\epsilon + \mathcal{O}\left(\epsilon^2\right) \,, \qquad n \ge  5 \,,
\end{equation}
are reproduced in \eqref{eq:Phi6final} by taking $q_{||,\ell}=\delta_{\ell,2}+\delta_{\ell,3}$.
\end{itemize}
In general, the operator construction parallels the one for the $\phi^4$ model in Sec.~\ref{sec: perturbativeSemiclassics}, where now one has to consider the spin $\ell$ representations of $SO(2,1)$.

\section{Outlook} \label{outlook}

Computing the scaling dimensions of neutral composite operators built from an arbitrarily large number of constituent fields is a formidable challenge for current methodologies. Here we showed that our semiclassical framework meets this challenge and provides  explicit, controlled solutions.  At the core of our method is a semiclassical implementation of the state--operator correspondence: scaling dimensions in flat space are obtained by quantizing the theory on the cylinder $\mathbb{R}\times S^{d-1}$, where scaling dimensions are identified with the corresponding energy levels on the cylinder (in units of the radius). Within this setting, we focus on the sector relevant to neutral composite operators and show that the problem reduces to determining the semiclassical energy spectrum associated with periodic, spatially homogeneous classical field configurations. A central result is that subleading semiclassical corrections admit a clean and geometric interpretation: they originate from deformations of the underlying periodic orbits. We illustrate the framework through two representative interacting fixed points, first  or $\phi^4$ theory near four dimensions and then for $\phi^6$ theory near three dimensions. In both cases, we semiclassically determine the full spectrum of operators transforming in traceless symmetric Lorentz representations, presenting the derivations and intermediate steps in a pedagogical manner.

The resulting methodology is conceptually transparent, computationally systematic, and readily portable: once the appropriate periodic classical solutions are identified, the same quantization machinery applies with minimal modification to a wider class of CFTs. 

Several natural extensions and complementary directions suggest themselves. An important avenue is to incorporate gauge and Yukawa interactions, enabling direct contact with Higgs-sector operators in the SMEFT and, more broadly, with weakly-coupled UV completions in which scalar, fermionic, and gauge degrees of freedom coexist. In the same spirit, it is of clear interest to extend the construction beyond purely scalar operators to include higher-spin sectors and operators built from fermions and gauge fields, thereby accessing a broader set of Lorentz representations and operator families.

Additionally, it would be interesting to employ numerical Monte Carlo simulations to access this sector of the CFT spectrum in the strongly coupled regime and investigate the generic $\displaystyle{\Delta_{n, q_\ell} \sim n^{\frac{d}{d-1}}}$ large $n$ behaviour, as recently done in \cite{Banerjee:2017fcx, Banerjee:2021bbw, Singh:2022akp, Cuomo:2023mxg, Hasenbusch:2025eno} for large charge operators.

 

\section*{Acknowledgments}
We would like to thank A. V. Bednyakov, J. Henriksson, and A. Miscioscia for their useful comments. We thank the authors of \cite{Bednyakov:2025usv} for sharing with us the preliminary results of their paper confirming our semiclassical computation for the $\phi^6$ theory. G.M. thanks the Simons Center for Geometry and Physics, Stony Brook University for the hospitality during the last stages of this work. The work of G.M. and F.S. is partially supported by the Carlsberg Foundation, semper ardens grant CF22-0922. The work of J.B. is supported by the Swiss National Science Foundation under grant number 200021\_219267.  

\newpage
\appendix

\section{Review of perturbative results in the $\phi^4$ model} \label{add1loop}

We collect here the known perturbative results for the anomalous dimension of composite operators in the $\phi^4$ theory that we use to compare with the semiclassical calculations presented in Sec.~\ref{sec: perturbativeSemiclassics}. These results have been determined in \cite{Henriksson:2022rnm, Henriksson:2025vyi, Kehrein:1994ff, Kehrein:1992fn, Derkachov:1997gc}. For the reader’s convenience, we also recall the values of $q_{1,\ell}$ and $q_{2,\ell}$ used to construct the operators from the semiclassical methodology. First, for Lorentz scalar $O(N)$-singlet operators of the form $\phi^n, \ n \geq 1$, we have
\begin{align}  \label{checkitout}
    q_{1,\ell} = \ q_{2,\ell}=0: \qquad \gamma_n &= \frac{n}{2(N+8)} (3n+N-4) \epsilon - \Bigg[\frac{17}{4(N+8)^2} n^3   
      \nonumber \\ &-\frac{604+(10-11 N)N}{4(N+8)^3}n^2   + \frac{576-N(118+35 N)}{4(N+8)^3}n\Bigg]\epsilon ^2  + \order{\epsilon^3} \ ,
\end{align}
and the scalar singlet operators with derivatives $\Box \phi^n, \ n =4,6,8,...$
\begin{align}\label{eq: known dev}
    q_{1,\ell} = 2 \delta_{\ell,1}, \ q_{2,\ell}=0: \qquad \gamma_{n,\{2\delta_{\ell,1}\}} = \frac{3 n^2+(N-12)n+8}{2(N+8)} \epsilon+ \mathcal{O}\left(\epsilon^2\right) \ ,
\end{align}
In the table below we display additional Ising CFT data.

\begin{table}[H]
\centering
\begin{tabular}{|c|l|l|}
\hline
Integers                                           & \multicolumn{1}{c|}{Operators}       & Anomalous dimension $\gamma_{n, q_\ell}$                                                                \\ \hline
$0$                                                & $\phi^n, \ n \geq 1$                 & $\frac{n(n-1)}{6} \epsilon - \frac{n(17n^2-67n+47)}{324} \epsilon^2 + \mathcal{O}\left(\epsilon^3\right)$ \\ \hline
$\delta_{\ell,2} $                                 & $\partial^2 \phi^n, \ n \geq 2$      & $\frac{(n-2)(3n+1)}{18}\epsilon + \order{\epsilon^2}$                                     \\
$\delta_{\ell,3} $                                 & $\partial^3 \phi^n, \ n \geq 3$      & $\frac{n^2-2n-2}{6}\epsilon+ \order{\epsilon^2}$                                          \\
$\delta_{\ell,4} $                                 & $\partial^4 \phi^n, \ n \geq 2$      & $\frac{(n-2)(5n-1)}{30}\epsilon + \order{\epsilon^2}$                                     \\
$\delta_{\ell,5}$                                  & $\partial^5 \phi^n, \ n \geq 3$      & $\frac{3n^2-7n-2}{18}\epsilon+ \order{\epsilon^2}$                                        \\ \hline
\multirow{3}{*}{$2 \delta_{\ell,2}$}               & $\partial^4 \phi^n, \ n \geq 4$      & $\frac{3n^2-7n-12}{18}\epsilon+ \order{\epsilon^2}$                                       \\
                                                   & $\partial^2 \Box \phi^n, \ n \geq 4$ & $\frac{3n^2-7n-4}{18}\epsilon+ \order{\epsilon^2}$                                        \\
                                                   & $\Box^2 \phi^n, \ n \geq 4$          & $\frac{n(3n-7)}{18}\epsilon+ \order{\epsilon^2}$                                          \\ \hline
\multirow{3}{*}{$\delta_{\ell,2}+\delta_{\ell,3}$} & $\partial^5 \phi^n, \ n \geq 5$      & $\frac{3n^2-8n-20}{18}\epsilon+ \order{\epsilon^2}$                                       \\
                                                   & $\partial^3 \Box \phi^n, \ n \geq 5$ & $\frac{3n^2-8n-10}{18}\epsilon+ \order{\epsilon^2}$                                       \\
                                                   & $\partial \Box^2 \phi^n, \ n \geq 5$ & $\frac{3n^2-8n-4}{18}\epsilon+ \order{\epsilon^2}$                                        \\ \hline
\end{tabular} 
\caption{Anomalous dimension of various families of operators in the traceless-symmetric Lorentz representations in the Ising CFT. In the first column, we have put the values of $q_{2,\ell}$ used to construct the operators in the semiclassical formalism.} \label{tableII}
\end{table}

\section{Elliptic functions properties}\label{jacobiapp}
In this section we summarise a few of the useful Jacobi elliptic functions and integrals properties that were used in carrying out our computations. All the integrals and functions depend on a real parameter $m\in(0,1)$ called the modulus. Firstly, the \emph{incomplete elliptic integrals of the first, second, and third kind} are defined as
\begin{align}
    F(\varphi|m) &\equiv  \int_0^{\sin \varphi} \frac{d t}{\sqrt{(1-mt^2)(1-t^2)}},\\
    E(\varphi|m) &\equiv \int_0^{\sin \varphi} dt ~ \sqrt{\frac{1-mt^2}{1-t^2}},\\
    \Pi(n|\varphi,m) &\equiv \int_0^{\sin \varphi} \frac{dt}{(1-nt^2) \sqrt{(1-t^2)(1-mt^2)}} ,
\end{align}
with the corresponding \textit{complete} elliptic integrals found by setting $\varphi=\pi/2$: 
\be
\mathbb{K}(m) \equiv F(\pi/2|m)\,, \quad \ \ELE(m) \equiv E(\pi/2|m)\,, \quad \ \Pi(n|m) \equiv\Pi(n|\pi/2,m)
\ee

Secondly, we introduce the \emph{Jacobi elliptic functions} $\sn,\cn,\dn$  defined by
\be
\cn(u\,|\,m)=\cos \varphi,~~~~~~~\sn(u\,|\,m)=\sin \varphi,~~~~~~\dn(u\,|\,m)=(1-m\,\sin^2 \varphi)^{1/2},
\ee 
where $\varphi$ is the \emph{Jacobi amplitude} 
\be
\varphi={\rm am} (u\,|\,m),~~~~{\rm with}~~~~u=\int_0^\varphi d\theta\  (1-m\,\sin^2\theta)^{-1/2}.
\ee
defined as the inverse of the incomplete elliptic integral of the first kind, i.e., as the solution of $F(\varphi|m) = u$.
The elliptic functions are doubly periodic and their period lies in the complex plane of $u$. Each has a real period and a complex one. We report here the real ones, which are $2\KK$ for $\dn$ and $4\KK$ for $\sn$ and $\cn$. It is worth mentioning that these functions have poles in the complex plane for imaginary values of $u$ at $u=i\KK(1-m), \ 3i\KK(1-m)$. \\

Next we introduce the Jacobi $\vartheta$ functions
\begin{align}
    &\vartheta_1(z,q)=\sum_{n=-\infty}^{\infty}(-1)^{n-1/2}q^{(n+1/2)^2}e^{(2n+1)iz},\\
    &\vartheta_4(z,q)= \sum_{n=-\infty}^{\infty} (-1)^n q^{n^2} e^{2n i z} .
\end{align}
The Jacobi $H$, $\Theta$ and $Z$ functions can then be defined as
\be\label{jacobidef}
H(u\,|\,m) = \vartheta_1\left(\frac{\pi\,u}{2\,\KK(m)}, q\right) \,, \qquad
\Theta(u\,|\,m) = \vartheta_4\left(\frac{\pi\,u}{2\,\KK(m)}, q\right) \,,\qquad
Z(u\,|\,m) = \frac{\pi}{2\,\KK(m)}\,\frac{\vartheta_4'(\frac{\pi\,u}{2\,\KK(m)}, q)}
{\vartheta_4(\frac{\pi\,u}{2\,\KK(m)}, q)}\,,
\ee
where the \emph{nome} $q$ is defined as
\be
q (m) =  \exp\left(-\pi\frac{\KK(1-m)}{\KK(m)}\right) \,.
\ee
Some useful periodicity properties are
\ba\label{periodicityH}
H(u+2\KK(m) \,|\,m)&=&-H(u \,|\,m) \,,\\
\Theta(u+2\KK(m) \,|\,m)&=&\Theta(u \,|\,m)\,,\\
Z(u+2\KK(m) \,|\,m)&=&Z(u\,|\,m) \,.
\ea
Lastly, using $Z(u|m) = E(u|m)-\ELE(m)/\KK(m) \ F(u|m)$, one can obtain the useful integral representation
\begin{equation}\label{intzeta}
Z(\sn^{-1}(u|m)\,|\,m)=\int_0^u\mathrm{d}t\left[\sqrt{\frac{1-mt^2}{1-t^2}}-\frac{\ELE(m)}{\KK(m)}
 \frac{1}{\sqrt{(1-t^2)(1-mt^2)}}\right] .
\end{equation}
from which Eq.~\eqref{useful} can be derived \cite{Beccaria:2010zn}.

\section{Small $\lambda n$ expansion of the stability angles} 
This section presents the perturbative computation of the stability angles for both the $\phi^4$ theory in $d = 4 - \epsilon$ and the $\phi^6$ theory in $d = 3 - \epsilon$. When an exact expression for the Bloch solutions of a differential operator with a periodic potential with period $\T$, is not available, we rely on perturbation theory. While this is not strictly necessary for the $\phi^4$ theory, for which we have an exact solution, this is not the case for the $\phi^6$ theory, where we have to resort to a perturbative approach. The strategy is standard: we expand both the differential operator and its eigenfunctions in the small $\lambda n$ limit, then solve the resulting eigenvalue equation order by order (see, for example, \cite{Beccaria:2010zn}). To this end, it is convenient to first perform a change of variables such that the period becomes independent on $\lambda n$ and equal to e.g., $2\pi$. Once the perturbative $2\pi$-periodic Bloch solution $\xi_{\ell,\pm}$ is determined, the corresponding stability angle $\nu_\ell$ is obtained as
\begin{align}
    \nu_\ell = \mp i \log \left( \frac{\xi_{\ell,\pm}(2\pi)}{\xi_{\ell,\pm}(0)} \right) \,,
\end{align}
up to the unphysical $2 \pi$ periodicity of $\nu_\ell$. Without further ado, we now present the details for each model and its associated quadratic fluctuation operators.\\

\subsection{$\phi^4$ theory}\label{smallnu}
As can be seen in Eq.~\eqref{mla}, the small $\lambda n$ expansion is equivalent to the small $m$ expansion of the stability angles. Hence, we expand the $L_\kappa$, $\kappa=1, \ 2$, quadratic fluctuation operators appearing in Eq.~\eqref{eq: lame_operator}, which correspond, respectively, to the $1-$gap and $2-$gap Lam\'e operators.\\
We start with the $2-$gap Lam\'e operator. It is useful to change the variable as
\be
z = \frac{2\pi}{4 \KK(m)}y \,,
\ee
such that the period of the potential, and hence of the Bloch solution, is $2 \pi$. Then the Lam\'e operator $L_2$ becomes
\begin{align}\label{eq:O2}
 L_2 &=   -\frac{\pi ^2}{4 \KK(m)^2}\partial _y^2+6m \ \sn\left(\left.\frac{2 \KK(m) }{\pi }y\right|m\right)^2-6 m-A_{\ell} (1-2 m) = -\partial_y^2 - A_{\ell}  \nonumber \\ &+\left(\frac{1}{2}\partial_y^2+ 2 A_{\ell}+6 \sin ^2 y-6\right) m + \frac{3}{32} (\partial_y^2-4 \cos (4 y)+4)  m^2  +\order{m^3} \ ,
\end{align}
with $A_{\ell}$ in Eq.~\eqref{lambda12}. We then perturbatively solve $L_2 \xi_{\ell,+} = 0$ by assuming the following ansatz  
\begin{equation}
   \xi_{\ell,+}(y)= e^{i y \left(p_0 + p_1 m + p_2 m^2  + \order{m^3}\right) } \left(\chi_0(y) + \chi_1(y) m + \chi_2(y) m^2  + \order{m^3}\right) \ , 
\end{equation}
with $\chi_j(y)= \chi_j(y+ 2\pi) \,, \  j=0,1,2$ such that $\xi_{\ell,+}$ corresponds to the Bloch solution with positive stability angles.
We find
\begin{align}
    \chi_0(y)&=1 \ , \\
        \chi_1(y)&=\frac{3i A_{\ell} \sin (2 y)-3 \sqrt{A_{\ell}} \cos (2 y)}{4 (A_{\ell}-1) \sqrt{A_{\ell}}} \ , \\
        \chi_2(y)&=\frac{1}{64 (A_{\ell}-4) (A_{\ell}-1)^2\sqrt{A_{\ell}}}\Big[3 (i (4 (A_{\ell}-4) (5 (A_{\ell}-1) A_{\ell}-6) \sin (2 y)\nonumber\\
        &+(A_{\ell}-10) (A_{\ell}-1) A_{\ell} \sin (4 y))-8 (A_{\ell}-4) \sqrt{A} (4 A_{\ell}-7) \cos (2 y)\nonumber\\
        &+(A_{\ell}-1) \sqrt{A_{\ell}} (A_{\ell}+8) \cos (4 y))\Big] \ , 
\end{align}
and
\begin{equation}
    \begin{split}
        p_0&=\sqrt{A_{\ell}} \ , \qquad p_1=-\frac{3 (A_{\ell}-2)}{4 \sqrt{A_{\ell}}} \ , \qquad p_2=-\frac{3 (A_{\ell} (A_{\ell} (13 A_{\ell}-49)+72)-24)}{64 (A_{\ell}-1) A_{\ell}^{3/2}} \,. 
    \end{split}
\end{equation}
We notice that the zero ($\ell=0$) and descendant ($\ell =1$) modes give rise to divergencies in the perturbative Bloch solution for $d=4$, where we have $A_0 = 1$ and $A_1 =4$. Therefore, these modes require separate treatment, as shown in Eq.~\eqref{finalform}. In particular, the small $m$ expansion of $\nu_{2,\ell=1}$ can also be found by first setting $\ell=1$ in the $L_2$ operator in Eq.~\eqref{eq:O2} and proceeding as described above. For $\ell>1$ we obtain
\begin{align}
  \nu_{2,\ell} &=-2\pi-i \log \frac{\xi_{\ell,+}(2 \pi)}{\xi_{\ell,+}(0)}=2\pi \left((p_0-1)+p_1 m+ p_2 m^2+\order{m^3}\right) = \nonumber \\ & 2 \pi  (\sqrt{A_{\ell}}-1)-\frac{3 \pi  (A_{\ell}-2) }{2 \sqrt{A_{\ell}}} m  -\frac{3 \pi  \left(13 A_{\ell}^3-49 A_{\ell}^2+72 A_{\ell}-24\right)}{32 (A_{\ell}-1) A_\ell^{3/2}} m^2  + \order{m^3} \,.
\end{align}

By proceeding in the same way for the $1-$gap Lamé operator $L_1$, for $\ell \ge 1$ we find
\begin{align}
    \nu_{1,\ell} &=2 \pi  \left(\sqrt{A_\ell}-1\right)-\frac{\pi  (3A_\ell-2) m}{2 \sqrt{A_\ell}} -\frac{\pi  (39 A_\ell^3-75 A_\ell^2+48 A_\ell-8) m^2}{32 (A_\ell-1) A_\ell^{3/2}} + \order{m^3} \ .
\end{align}
Armed with the perturbative expansion of the stability angles, it is then straightforward to obtain the results in equations \eqref{lasterm} and \eqref{lastterm2} for the regularized sums over $\ell$ appearing in our final expression for $\Delta_{n,q_{\ell}}$ \eqref{eq: excited Delta ON}.

\subsection{$\phi^6$ theory}\label{smallnu6}
We can now proceed with the computation of the stability angles for the $O(N)$ $\phi^6$ model. We start with the quadratic fluctuation operator $  \mathcal{O}_{||,\ell}$ given in Eq.~\eqref{eq:quad_op_6}. Using the solution \eqref{phi6fullsol} and expanding for $\lambda n\ll1$, this becomes

\begin{equation}\label{eq:quad_op_6app}
\begin{split}
  \mathcal{O}_{||,\ell}  &=-\left(\partial_y^2+A_{\ell}\right)-\frac{5}{2\pi^2}(\partial_y^2+8\cos^4y)(\lambda n)^2\\
     &+\frac{5}{96\pi^4}\left[101\partial_y^2-32\cos^4y(-37+14\cos(2y)+\cos(4y)) \right](\lambda n)^4 + \order{\lambda^6 n^6} \ ,
\end{split}
\end{equation}
where we introduced the dimensionless variable $\displaystyle{y=\frac{2\pi}{\T}t}$. Again, our ansatz for the Bloch solutions of the $\mathcal{O}_{||,\ell}$ operator is
\begin{align}
  &\xi_{\ell,+}(y)=e^{i y \left(p_{0} + p_{1} (\lambda n)^2 + p_{2}  (\lambda n)^4 + \order{ (\lambda n)^6}\right) } \left(\chi_{0}(y) + \chi_{1}(y)  (\lambda n)^2 + \chi_{2}(y)  (\lambda n)^4  + \order{ (\lambda n)^6}\right)  \,,
\end{align}
where, with a slight abuse of notation, we employ the same names as for the $\phi^4$ model and the $\chi$'s are $2\pi$-periodic functions.
We find
\begin{equation}
    \chi_{0}(y)=1 \ ,
\end{equation}
\begin{equation}
    \begin{split}
         \chi_{1}(y)&=\frac{i}{16 (-4 + A_{\ell}) (-1 + A_{\ell}) \sqrt{A_{\ell}} \pi^2}\Bigg(-4 (-4 + A_{\ell}) (-1 + A_{\ell}) (-15 + 5 A_{\ell} + 4  \sqrt{A_{\ell}} p_1 \pi^2) y\\
         &+ 
 5 (8i (-4 + A_{\ell}) \sqrt{A_{\ell}} \cos{2 y} + 2i (-1 + A_{\ell}) \sqrt{A_{\ell}} \cos{4 y} + 
    8 (-4 + A_{\ell}) A_{\ell} \sin{2 y}\\
    &+ (A_{\ell}-1) A_{\ell} \sin{4 y})\Bigg) \ ,
    \end{split}
\end{equation}
\begin{equation}
\begin{split}
   \chi_{2}(y)&=\frac{i}{3072 \pi ^4 (A_{\ell}-16) (A_{\ell}-9) (A_{\ell}-4)^2 (A_{\ell}-1)^2 A_{\ell}^{3/2}}\Bigg(-80 i (A_{\ell}-16) (A_{\ell}-9) (A_{\ell}\\
   &-4) (A_{\ell} (179 A_{\ell}-325)-1294) A_{\ell}^{3/2} \cos (2 y)+16 (A_{\ell}-16) (A_{\ell}-9) (A_{\ell}-4) (A_{\ell}\\
   &-1) y \left(A_{\ell} (A_{\ell} (-192 \pi ^4 A_{\ell}^{3/2} p_2+955 A_{\ell}^2+960 \pi ^4
   \sqrt{A_{\ell}} p_2-7325 A_{\ell}+13945\right)\\
   &-768 \pi ^4 \sqrt{A_{\ell}} p_2+1425)-5400)-960 i (A_{\ell}-16) (A_{\ell}-9) (A_{\ell}-1) (6 A_{\ell}^2-20 A_{\ell}\\
   &-31) A_{\ell}^{3/2} \cos (4 y)+5 A_{\ell}
   (2 (A_{\ell}-1) ((A_{\ell}-4) A_{\ell} (8 (A_{\ell}-16) (A_{\ell} (6 A_{\ell}-95)+209) \sin (6 y)\\
   &+(A_{\ell}-9) (A_{\ell}-1) (2 A_{\ell}-53) \sin (8 y))-24 (A_{\ell}-16) (A_{\ell}-9) (A_{\ell}+4) (7 (A_{\ell}-4) A_{\ell}\\
   &-15) \sin (4 y))-16 (A_{\ell}-16) (A_{\ell}-9) (A_{\ell}-4)
   ((A_{\ell}-1) A_{\ell} (314 A_{\ell}-1091)-1440) \sin (2 y)\\
   &+i (A_{\ell}-184) (A_{\ell}-9) (A_{\ell}-4) \sqrt{A_{\ell}} (A_{\ell}-1)^2 \cos (8 y)+48 i (A_{\ell}-16) (A_{\ell}-4) \sqrt{A_{\ell}} ((A_{\ell}\\
   &-35) A_{\ell}+74) (A_{\ell}-1) \cos (6 y))\Bigg) \ ,
\end{split}
\end{equation}
while for the $p_{i}$'s we find
\begin{align}
    p_{0} &= \sqrt{A_{\ell}} \,, \qquad 
   p_{1}=\frac{5 (3 - A_{\ell})}{4 \sqrt{A_{\ell}} \pi^2}\,,\nonumber\\
    p_{2}&=\frac{-5400 + 5 A_{\ell} (285 + A_{\ell} (2789 + A_{\ell} (-1465 + 191 A_{\ell})))}{192 (-4 + A_{\ell}) (-1 + A_{\ell}) A_{\ell}^{3/2} \pi^4} \,,
\end{align}

\noindent
For $\ell > 1$, one can then obtain the stability angles as
\begin{align}
    \nu_{||,\ell}&=-2\pi-i\log\Big\{\frac{\xi_{\ell,+}(2\pi)}{\xi_{\ell,+}(0)}\Big\} =2 (-1 + \sqrt{A_{\ell}}) \pi +(\lambda n)^2\frac{5 (3 - A_{\ell})}{2 \sqrt{A_{\ell}} \pi} \\
    &+(\lambda n)^4\frac{-5400 + 5 A_{\ell} (285 + A_{\ell} (2789 + A_{\ell} (-1465 + 191 A_{\ell})))}{96 (-4 + A_{\ell}) (-1 + A_{\ell}) A_{\ell}^{3/2} \pi^3}+ \order{(\lambda n)^{6}} \,.\nonumber
\end{align}

As happened for the $\phi^4$ case, the perturbative Bloch solution for the operator $ \mathcal{O}_{||,\ell} $ is singular for $\ell=0$, corresponding to the zero mode stemming from time-translation symmetry, and for $\ell=1$. To obtain the $\ell=1$ stability angles we then proceed by by directly plugging $\ell=1$ in the differential operator\footnote{The $\ell=0, \ 1$ modes will be treated separately in the sum in Eq.~\eqref{eq:phi6excited}, and hence can be directly computed in $d=3$.} \eqref{eq:quad_op_6app} for $d=3$, finding 
\begin{equation}
    \nu_{||,1}=\T \ .
\end{equation}
Therefore, the $\nu_{||,1}$ corresponds to the mode generating descendant states. We can then turn our attention to the fluctuation operator for the transversal modes $\tilde{\phi}_a$ ($a=2,\dots,N$) in \eqref{eq:quad_op_6_2}. The procedure being the same, we leave here directly the result for the stability angles $\nu_{\perp,\ell}$ for $\ell > 1$:
{\small
\begin{equation}
    \begin{split}
        \nu_{\perp,\ell}&=2 (\sqrt{A_{\ell}}-1) \pi +\frac{3 - 5 A_{\ell}}{2 \sqrt{A_{\ell}} \pi}(\lambda n)^2 +\frac{(\lambda n)^4}{96 (-4 + A_{\ell}) (-1 + A_{\ell}) A_{\ell}^{3/2} \pi^3} \Big(216 + 5 A_{\ell} (-93  \\ & +102 (-4 + A_{\ell}) (-1 + A_{\ell}) + 21 A_{\ell} - 
    191 (-4 + A_{\ell}) (-1 + A_{\ell}) A_{\ell})\Big)+\order{(\lambda n)^{6}} \ .
    \end{split}
\end{equation}}\normalsize
As for the $\phi^4$ model we observe that $\nu_{\perp,0} = 0$, corresponding to the expected $N-1$ zero modes arising from the spontaneous symmetry breaking $O(N) \to O(N-1)$ induced by the classical solutions. Finally, we note that the Bloch solutions for the $\cO_{\perp,\ell}$ operators are singular also for $\ell=1$ even if the latter is not the descendant mode. By again replacing $\ell=1$ in the $d=3$ operator before determining its Bloch solutions, we find 
\begin{equation}
    \nu_{\perp,1}= 4\pi-\frac{7}{\pi}(\lambda n)^2+\frac{6727 }{240\pi^3}(\lambda n)^4 + \order{(\lambda n)^6} \ .
\end{equation}

\section{Explicit $8$-loop results} \label{8loop}

In this section, we provide explicit $8$-loop results for the scaling dimension $\Delta_{n,q_\ell}$. Moreover, to ease the comparison to perturbative calculations, we also report the corresponding anomalous dimensions $\gamma_{n,q_\ell}$ as a function of the couplings to the $4$-loop order. For the scaling dimensions $\Delta_{n,q_\ell}$, we use the notation of equations \eqref{roperto} and \eqref{cuoppo}, and present the value of the $c_{ik}$ coefficients. For the $\phi^4$ theory studied in Sec.~\ref{phi4} we obtain

\allowdisplaybreaks

\begin{align}
    \gamma_{n,q_\ell} &= \bigg[\frac{3 n^2}{16 \pi ^2} + n \bigg( \frac{N-4}{16 \pi^2}  -\sum_{\ell=1}^{\infty} \frac{2(3 \ell+3) q_{1,\ell} + 6(\ell-1) q_{2,\ell}}{16 \pi ^2 (\ell+1)}\bigg)\bigg] \lambda \nonumber\\
    &+\bigg[-\frac{17 n^3}{256 \pi ^4} + n^2\bigg(-\frac{11 n^2 (N-4)}{256 \pi ^4}\nonumber\\ 
    &+ \sum_{\ell=1}^\infty \frac{q_{1,\ell}}{256 \pi ^4 \ell (\ell+1)^3 (\ell+2)} \bigg(-4+30\ell+225 \ell^2 + 363 \ell^3 + 231 \ell^4 + 51 \ell^5\bigg)   \nonumber\\
    &+ \sum_{\ell=1}^\infty \frac{q_{2,\ell}}{256 \pi ^4 \ell (\ell+1)^3 (\ell+2)}\bigg(-36 - 258 \ell - 111 \ell^2 + 171 \ell^3 + 183 \ell^4 + 51 \ell^5\bigg)\bigg)\bigg] \lambda^2\nonumber\\
    &+ \bigg[\frac{375 n^4}{8192 \pi ^6} + n^3 \bigg(\frac{ (32 N \zeta (3)+285 N+832 \zeta (3)-1992)}{8192 \pi ^6} \nonumber\\
    &+ \sum_{\ell=1}^\infty \frac{q_{1,\ell}}{2048 \pi ^6 \ell^2 (\ell+1)^5 (\ell+2)^2} \bigg(-8 + 24 \ell - 392 \ell^2 - 4640 \ell^3 - 14985 \ell^4\nonumber\\
    &- 23987 \ell^5 - 21762 \ell^6 - 
 11390 \ell^7 - 3205 \ell^8 - 375 \ell^9\bigg)\nonumber\\
    &+ \sum_{\ell=1}^\infty \frac{q_{2,\ell}}{2048 \pi ^6 \ell^2 (\ell+1)^5 (\ell+2)^2} \bigg(216 + 1656 \ell + 7992 \ell^2 + 12720 \ell^3 + 7035 \ell^4\nonumber\\
    &- 5763 \ell^5 - 
 12378 \ell^6 - 8670 \ell^7 - 2865 \ell^8 - 375 \ell^9\bigg)  \bigg)\bigg] \lambda^3\nonumber\\
 &+ \bigg[-\frac{10689 n^5}{262144 \pi ^8} + n^4 \bigg(\frac{1}{262144 \pi ^8} \bigg(-8 (125 N+2224) \zeta (3)\nonumber\\
 &-320 (N+80) \zeta (5)-9975 N+81732\bigg)\nonumber\\
 &+ \sum_{\ell=1}^\infty \frac{q_{1,\ell}}{262144 \pi ^8 \ell^3 (\ell+1)^7 (\ell+2)^3} \bigg(-320 + 480 \ell - 5360 \ell^2 + 99080 \ell^3\nonumber\\
 &+ 1632380 \ell^4 + 7922330 \ell^5 + 
 20400275 \ell^6 + 32673605 \ell^7 + 34676695 \ell^8 + 24905085 \ell^9\nonumber\\
 &+ 
 12000585 \ell^{10} + 3720375 \ell^{11} + 670785 \ell^{12} + 53445 \ell^{13}\bigg) \nonumber\\
 &+ \sum_{\ell=1}^\infty \frac{q_{2,\ell}}{262144 \pi ^8 (\ell-1) \ell^3 (\ell+1)^7 (\ell+2)^3 (\ell+3)} \bigg(77760 + 702432 \ell + 3186000 \ell^2\nonumber\\
 &+ 11758824 \ell^3 + 22248876 \ell^4 + 
 18544050 \ell^5 - 10462665 \ell^6 - 44928015 \ell^7 - 47217960 \ell^8\nonumber\\
 &- 
 14195460 \ell^9 + 17511990 \ell^{10} + 23441130 \ell^{11} + 13324980 \ell^{12} + 
 4229610 \ell^{13}\nonumber\\
 &+ 729675 \ell^{14} + 53445 \ell^{15}\bigg)\bigg)\bigg] \lambda^4 +\cO(\lambda^5, \lambda^4 n^3, \lambda^3 n^2, \lambda^2 n , \lambda n^0) \ ,
\end{align}
and
\begin{align}
    c_{00} &= 1 \ , \qquad   c_{01} = \frac{3}{2 (N+8)} \ , \qquad    c_{02} = -\frac{17}{4 (N+8)^2}\ , 
\end{align}
\begin{equation}
     c_{03} = \frac{375 }{16 (N+8)^3} \ , \qquad  c_{04} = -\frac{10689}{64 (N+8)^4} \ , \qquad  c_{05} = \frac{87549}{64 (N+8)^5} \ ,
\end{equation}

\begin{equation}
   c_{06} = -\frac{3132399}{256 (N+8)^6} \ , \qquad  c_{07} = \frac{238225977}{2048 (N+8)^7} \ , \qquad    c_{08} = -\frac{18945961925}{16384 (N+8)^8} \ , 
\end{equation}

\begin{equation}
     c_{10} = \sum_{\ell=1}^\infty (q_{1,\ell}+q_{2,\ell}) \ell \ , 
\end{equation}

\begin{align}
    c_{11} &= -\frac{1}{2} +\frac{ (N-4)}{2 (N+8)} - \sum_{\ell=1}^\infty \bigg[q_{1,\ell}\frac{(3 \ell+1) n}{(\ell+1) (N+8)} + q_{2,\ell}\frac{3 (\ell-1) n}{(\ell+1) (N+8)} \bigg] \ .
\end{align}

\begin{align}
    c_{12} &= \frac{1}{4 (N+8)^3} \bigg[-11 N^2+10 N+604\bigg]\nonumber\\
    &+\sum_{\ell=1}^\infty \frac{q_{1,\ell}}{4 \ell (\ell+1)^3 (\ell+2) (N+8)^2} \bigg[-4 + 30 \ell +225 \ell^2+363 \ell^3 +231 \ell^4+51 \ell^5\bigg]\nonumber\\
    &+\sum_{\ell=1}^\infty \frac{q_{2,\ell}}{4 \ell (\ell+1)^3 (\ell+2) (N+8)^2} \bigg[-36-258 \ell-111 \ell^2+171 \ell^3+183 \ell^4+51 \ell^5\bigg] \ .
\end{align}

\begin{align}
    c_{13} &= \frac{1}{16 (N+8)^4}\bigg[(N+8) (32 N \zeta (3)+285 N+832 \zeta (3)-1992)-408 (3 N+14)\bigg]\nonumber\\
    &+\sum_{\ell=1}^\infty \frac{q_{1,\ell}}{4 \ell^2 (\ell+1)^5 (\ell+2)^2 (N+8)^3} \bigg[-8+24\ell -  392 \ell^2-4640 \ell^3-14985 \ell^4-23987 \ell^5\nonumber\\
    &-21762 \ell^6-11390 \ell^7-3205 \ell^8 -375 \ell^9\bigg] \nonumber\\
    &+\sum_{\ell=1}^\infty \frac{q_{2,\ell}}{4 \ell^2 (\ell+1)^5 (\ell+2)^2 (N+8)^3} \bigg[216+1656\ell +7992 \ell^2+12720 \ell^3+7035 \ell^4-5763 \ell^5\nonumber\\
    &-12378 \ell^6 -8670 \ell^7 -2865 \ell^8-375 \ell^9\bigg] \ .
\end{align}

\begin{align}
    c_{14} &= \frac{1}{64 (N+8)^6} \bigg[(N+8)^2 (-8 (125 N+2224) \zeta (3)-320 (N+80) \zeta (5)\nonumber\\
    &-9975 N+81732)+13500 (3 N+14) (N+8)\bigg]\nonumber\\
    &+ \sum_{\ell=1}^\infty \frac{q_{1,\ell}}{64 \ell^3 (\ell+1)^7 (\ell+2)^3 (N+8)^4} \bigg[ -320 +480 \ell -5360 \ell^2 +99080 \ell^3 +1632380 \ell^4   \nonumber\\
    &+7922330 \ell^5 +20400275 \ell^6 +32673605 \ell^7 +34676695 \ell^8 +24905085 \ell^9 +12000585 \ell^{10}\nonumber\\
    &+3720375 \ell^{11} +670785 \ell^{12} +53445 \ell^{13}  \bigg]\nonumber\\
    &+ \sum_{\ell=1}^\infty \frac{q_{2,\ell}}{64 (\ell-1) \ell^3 (\ell+1)^7 (\ell+2)^3 (\ell+3) (N+8)^4} \bigg[77760+702432 \ell +3186000 \ell^2\nonumber\\
    &+11758824 \ell^3+22248876 \ell^4+18544050 \ell^5-10462665 \ell^6-44928015 \ell^7 -47217960 \ell^8\nonumber\\
    &-14195460 \ell^9 +17511990 \ell^{10}+23441130 \ell^{11}+13324980 \ell^{12}+4229610 \ell^{13}+729675 \ell^{14}\nonumber\\
    &+53445 \ell^{15}\bigg] \ .
\end{align}

\begin{align}
    c_{15} &= \frac{1}{256 (N+8)^7} \bigg[-513072 (N+8) (3 N+14) + (N+8)^2\bigg(390469 N-3466540\nonumber\\
    &+64 (569 N+8422) \zeta (3)+2560 (5 N+238) \zeta (5)+3584 (N+242) \zeta (7))\bigg)\bigg]\nonumber\\
    &+ \sum_{\ell=1}^\infty \frac{q_{1,\ell}}{32 \ell^4 (\ell+1)^9 (\ell+2)^4 (N+8)^5} \bigg[-448+192 \ell -4800 \ell^2+37440 \ell^3 -877560 \ell^4\nonumber\\
    &-18651624 \ell^5-120982164 \ell^6 -430134792 \ell^7 -989638380 \ell^8 -1587224247 \ell^9\nonumber\\
    &-1842129912 \ell^{10} -1573697076 \ell^{11} -992085696 \ell^{12}-456505602 \ell^{13}-149159220 \ell^{14}\nonumber\\
    &-32788140 \ell^{15} -4347420 \ell^{16} -262647 \ell^{17}\bigg]\nonumber\\
    &+ \sum_{\ell=1}^\infty \frac{q_{2,\ell}}{32 (\ell-1)^2 \ell^4 (\ell+1)^9 (\ell+2)^4 (\ell+3)^2 (N+8)^5} \bigg[979776+10404288 \ell +53265600 \ell^2\nonumber\\
    &+183643200 \ell^3 +565017768 \ell^4  +1166712984 \ell^5+1317304476 \ell^6 -168148152 l^7\nonumber\\
    &-3298922118 \ell^8 -5258885247 \ell^9-2954942100 \ell^{10} + 2228558778 \ell^{11} +5309807490 \ell^{12}\nonumber\\
    &+3923943471 \ell^{13}+601136676 \ell^{14} -1364272812 \ell^{15} -1396309698 \ell^{16}  -714318129 \ell^{17}\nonumber\\
    &-226265004 \ell^{18} -44949366 \ell^{19} -5162850 \ell^{20} -262647 \ell^{21}\bigg] \ .
\end{align}

\begin{align}
    c_{16} &= \frac{1}{1024 (N+8)^9} \bigg[21011760 (N+8)^2 (3 N+14) +(8+N)^3 \bigg(152671980\nonumber\\
    &-16277877 N-144 (10045 N+128456) \zeta (3)-64 (8141 N+276088) \zeta (5)\nonumber\\
    &-64 (8141 N+276088) \zeta (5)-43008 (N+728) \zeta (9)\bigg)\bigg]\nonumber\\
    &+ \sum_{\ell=1}^{\infty} \frac{q_{1,\ell}}{256 l^5 (\ell+1)^{11} (\ell+2)^5 (N+8)^6} \bigg[-10752+5376 \ell -94080 \ell^2
    +460544 \ell^3 -4783296 \ell^4\nonumber\\
    &+133975072 \ell^5+3499108592 \ell^6+28437827936 \ell^7+129103512720 \ell^8 +387706108258 \ell^9\nonumber\\
    & +833242055639 \ell^{10}+1338020345473 \ell^{11}+1645624713761 \ell^{12}+1571035134081 \ell^{13}\nonumber\\
    &+1170068003734 \ell^{14} +678301608026 \ell^{15}+303055967858 \ell^{16}+102344918340 l^{17}\nonumber\\
    &+25262072115 \ell^{18} +4298882805 \ell^{19}+450657165 \ell^{20}+21926793 \ell^{21}\bigg]\nonumber\nonumber\\
    &+\sum_{\ell=1}^\infty \frac{q_{2,\ell}}{256 \ell^5 (\ell+1)^{11} (\ell+2)^5 \left(\ell^2+2 \ell-3\right)^3 (N+8)^6} \bigg[211631616+2598365952 \ell+15194646144 \ell^2\nonumber\\
    &+57540471552 \ell^3+165907818432 \ell^4+446917370400 l^5+963061920432 \ell^6+1353199208544 \ell^7\nonumber\\
    &+416817352992 \ell^8-3011397970806 \ell^9-7173477472929 \ell^{10}-6727051591827 l^{11}+1184945360340 \ell^{12}\nonumber\\
    &+10675433891892 \ell^{13}+12083745684864 \ell^{14}+3873679897584 \ell^{15}-5453816763948 \ell^{16}\nonumber\\
    &-8053490451048 \ell^{17}-4524889379514 \ell^{18}-243005112798 \ell^{19} +1575351594012 \ell^{20}+1367880053484 \ell^{21}\nonumber\\
    &+658879918536 \ell^{22} +210338621976 \ell^{23}+45725193948 \ell^{24}+6558720798 \ell^{25}\nonumber\\
    &+562606947 \ell^{26} +21926793 \ell^{27}\bigg] .
\end{align}

\begin{align}
    c_{17} &= \frac{1}{2048 (N+8)^{10}} \bigg[(N+8)^3-451065456 (N+8)^2 (3 N+14)\bigg(-3450397870\nonumber\\
    &+353561278 N+1664 (18244 N+217871) \zeta (3)+128 (86101 N+2183276) \zeta (5)\bigg)\nonumber\\
    &+7168 (569 N+53134) \zeta (7)+258048 (5 N+2182) \zeta (9)+270336 (N+2186) \zeta (11)\bigg]\nonumber\\
    &+ \sum_{\ell=1}^\infty \frac{q_{1,\ell}}{256 \ell^6 (\ell+1)^{13} (\ell+2)^6 (N+8)^7} \bigg[-33792-46080 \ell-286720 \ell^2+845824 \ell^3\nonumber\\
    &-7108352 \ell^4 +82036480 \ell^5 -2693888768 \ell^6-83590377728 \ell^7 -816981002200 \ell^8\nonumber\\
    &-4515761357208 \ell^9-16742706266472 \ell^{10}-45144785017008 \ell^{11}-92673378522423 \ell^{12}\nonumber\\
    &-148960492378341 \ell^{13} -190816993106250 \ell^{14} -196880766385590 \ell^{15}-164488460944605 \ell^{16}\nonumber\\
    &-111366276254055 \ell^{17}-60882846055908 \ell^{18}-26650130167956 \ell^{19}-9206448795105 \ell^{20}\nonumber\\
    &-2453177049411 \ell^{21}-486251550666 \ell^{22} -67482399654 \ell^{23}-5849147859 \ell^{24}-238225977 \ell^{25}\bigg]\nonumber\\
    &+ \sum_{\ell=1}^\infty \frac{q_{2,\ell}}{256 (\ell-1)^4 \ell^6 (\ell+1)^{13} (\ell+2)^6 (\ell+3)^4 (N+8)^7} \bigg[5986151424+83624721408 \ell\nonumber\\
    &+554515144704 \ell^2+2348654082048 \ell^3+7293989461248 \ell^4+18498040803072 \ell^5\nonumber\\
    &+45095052665088 \ell^6 +98956064240640 \ell^7 +163184975252424 \ell^8 +116254107414216 \ell^9\nonumber\\
    &-273608196305832 \ell^{10}-1014982242128640 \ell^{11}-1412548620700491 \ell^{12}-433322782964685 \ell^{13}\nonumber\\
    &+1791611712438786 \ell^{14} +3264654858790518 \ell^{15} +2062019653248717 \ell^{16} -1059343986044133 \ell^{17}\nonumber\\
    &-3186456219137592 \ell^{18}-2568222018174744 \ell^{19}-340607083667382 \ell^{20}+1262761511412294 \ell^{21}\nonumber\\
    &+1351182817435212 \ell^{22} +613273517052516 \ell^{23}-14792402137302 \ell^{24}-220965613279434 \ell^{25}\nonumber\\
    &-172094931742896 \ell^{26}-79869716827992 \ell^{27}-25773346636191 \ell^{28}-5976448837209 \ell^{29}\nonumber\\
    &-985507288254 \ell^{30} -110318194026 \ell^{31} -7541952543 \ell^{32}-238225977 \ell^{33}\bigg] \ .
\end{align}
 
\begin{align}
    c_{18} &= \frac{1}{16384 (N+8)^{12}} \bigg[40021964136 (N+8)^3 (3 N+14)-3 (N+8)^4 (720 (1225603 N+13948688) \zeta (3)\nonumber\\
    &+10752 (29857 N+623570) \zeta (5)+384 (320629 N+20739776) \zeta (7)+57344 (757 N+207008) \zeta (9)\nonumber\\
    &+2523136 (5 N+6556) \zeta (11)+2342912 (N+6560) \zeta (13)+10543237491 N-106248256444)\bigg]\nonumber\\
    &+ \sum_{\ell=1}^\infty \frac{q_{1,\ell}}{16384 \ell^7 (\ell+1)^{15} (\ell+2)^7 (N+8)^8} \bigg[-7028736-15409152 \ell -65286144 l^2 +94402560 l^3\nonumber\\
    &-895564800 \ell^4 + 7226740224 \ell^5 - 95523641088 \ell^6 + 3602830972032 \ell^7+129594439892160 \ell^8\nonumber\\
    &+ 1480527905857440 \ell^9+9648848298914352 \ell^{10}+42588461257143912 \ell^{11} + 138199974769170060 \ell^{12}\nonumber\\
    &+345649691618678970 \ell^{13} + 686495818382521635 \ell^{14} + 
 1104311450889360621 \ell^{15}\nonumber\\ 
 &+ 1457929756931196267 \ell^{16}+1593292765922153469 \ell^{17} + 1448531180919721215 \ell^{18} \nonumber\\
 &+ 1097660067394927905 \ell^{19}+692655374577358863 \ell^{20} + 362588557852804437 \ell^{21}\nonumber\\
 &+156340961643454281 \ell^{22} + 54909731228876775 \ell^{23} + 
 15450074207481825 \ell^{24}\nonumber\\
 &+3397942116529191 \ell^{25} + 562520953332261 \ell^{26} + 65900444518755 \ell^{27}\nonumber\\
 &+4868663749785 \ell^{28} + 170513657325 \ell^{29}\bigg]\nonumber\\
 &+ \sum_{\ell=1}^\infty \frac{q_{2,\ell}}{16384 (\ell-1)^5 \ell^7 (\ell+1)^{15} (\ell+2)^7 (\ell+3)^5 (N+8)^8} \bigg[11206075465728 + 175705516597248 \ell \nonumber\\
 &+1307638861664256 \ell^2 + 6185234494494720 \ell^3 + 21132670825313280 \ell^4+56832574490408448 \ell^5\nonumber\\
 &+130874881999189248 \ell^6 + 295949706048825984 \ell^7+652737888327061440 \ell^8\nonumber\\
 &+ 1211586108766267680 \ell^9+1306653340175209584 \ell^{10} - 1117523574503232696 \ell^{11}\nonumber\\
 &-7947792383677769700 \ell^{12} - 15444843708018240270 \ell^{13}-11553285058938797625 l^{14}\nonumber\\
 &+12488770099185820425 \ell^{15} + 42451204991536810800 \ell^{16} + 
 44140549016597239980 \ell^{17}\nonumber\\
 &+1670513763607286490 \ell^{18} - 52567587488260666890 \ell^{19}-65887664004194934060 \ell^{20}\nonumber\\
 &-25126345345639866210 \ell^{21} + 28608874270888487085 \ell^{22} + 
 48463585400933587035 \ell^{23}\nonumber\\
 &+28926988170406331040 \ell^{24} - 1010388104137416888 \ell^{25}-15913758926781354708 \ell^{26}\nonumber\\
 &-13560748314075768276 \ell^{27} - 5182003976943180060 \ell^{28} + 
 471846795038742510 \ell^{29}\nonumber\\
 &+1964518364722196241 \ell^{30} + 1418071022918285391 \ell^{31}+642511511319423024 \ell^{32}\nonumber\\
 &+209740506581538540 \ell^{33} + 51108345083227050 \ell^{34} + 
 9297838437733494 \ell^{35}\nonumber\\
 &+1233141908806764 \ell^{36} + 113056267689090 \ell^{37} + 6421335697755 \ell^{38}+170513657325 \ell^{39}\bigg] \ .
\end{align}
Finally, for the $\phi^6$ theory analyzed in Sec.~\ref{phi6}, we obtain
\begin{equation}\label{eq:Phi6anomalousdim}
\begin{split}
& \gamma_{n,q_{\ell}}=\Bigg[\frac{5  }{6}n^3-\Bigg(\sum_{\ell=1}^{\infty}q_{||,\ell}\frac{5 (\ell-1) 
   }{(2 \ell+1) }+\sum_{\ell=1}^{\infty}q_{\perp,\ell}\frac{5 \ell+1  }{(2
   \ell+1)}-\frac{ N-6  }{2}\Bigg)n^2+\cO(n) \Bigg]\Bigg(\frac{\lambda^2}{4\pi^2}\Bigg)\\
   &-\Bigg[\frac{131 }{24 }n^5-\Bigg(\frac{ -1136 N-27 \pi ^2 (N+24)+9216 }{192}\\
   &+\sum_{\ell=1}^{\infty}q_{||,\ell}\frac{450 + 4680 \ell + 1445 \ell^2 - 17705 \ell^3 - 
 25150 \ell^4 + 2320 \ell^5 + 23480 \ell^6 + 
 10480 \ell^7}{12 \ell (2 \ell+1)^3 \left(4 (\ell+2) \ell^2+\ell-3\right)}\\
&+\sum_{\ell=1}^{\infty}q_{\perp,\ell}\frac{18 - 360 \ell - 3835 \ell^2 - 7625 \ell^3 + 
 6290 \ell^4 + 34000 \ell^5 + 34040 \ell^6 + 
 10480 \ell^7}{12 \ell (2 \ell+1)^3 \left(4 (\ell+2) \ell^2+\ell-3\right)}\Bigg)n^4\\
 &+\cO\left(n^3\right)\Bigg]\Bigg(\frac{\lambda^2}{4\pi^2}\Bigg)^2+\order{\Bigg(\frac{\lambda^2}{4\pi^2}\Bigg)^3} \ ,
\end{split}
\end{equation}
and
\begin{equation}
    c_{00}=\frac{1}{2} \ , \qquad     c_{01}=\frac{5}{6(3N+22)} \ ,
\end{equation}
\begin{equation}
    c_{02}=-\frac{131 }{24 (3 N+22)^2} \ , \qquad c_{03}=\frac{4915 }{72 (3 N+22)^3} \ , \qquad  c_{04}=-\frac{3817055}{3456 (22 + 3 N)^4} \ ,
\end{equation}
\begin{equation}
    c_{10}= \sum_{\ell=1}^{\infty}(q_{||,\ell}+q_{\perp,\ell})\ell \ , 
\end{equation}
\begin{equation}
\begin{split}
    c_{11}&=-\Bigg(\sum_{\ell=1}^{\infty}q_{||,\ell}\frac{5 (\ell-1) 
   }{(2 \ell+1) (3 N+22)}+\sum_{\ell=1}^{\infty}q_{\perp,\ell}\frac{5 \ell+1  }{(2
   \ell+1) (3 N+22)}-\frac{ N-6  }{6 N+44}\Bigg) \ ,
\end{split}
\end{equation}
\begin{equation}
\begin{split}
    c_{12}&=\Bigg(\frac{ -1136 N-27 \pi ^2 (N+24)+9216 }{192 (3 N+22)^2}\\
   &+\sum_{\ell=1}^{\infty}q_{||,\ell}\frac{10480 \ell ^7+23480 \ell ^6+2320 \ell ^5-25150 \ell ^4-17705 \ell ^3+1445 \ell ^2+4680
   \ell +450}{12 \ell (2 \ell+1)^3 \left(4 (\ell+2) \ell^2+\ell-3\right) (3 N+22)^2}\\
&+\sum_{\ell=1}^{\infty}q_{\perp,\ell}\frac{10480 \ell ^7+34040 \ell ^6+34000 \ell ^5+6290 \ell ^4-7625 \ell ^3-3835 \ell ^2-360 \ell
   +18}{12 \ell (2 \ell+1)^3 \left(4 (\ell+2) \ell^2+\ell-3\right) (3 N+22)^2}\Bigg) \ ,
\end{split}
\end{equation}
\begin{equation}
    \begin{split}
        c_{13}&=\frac{n^6 \left(567 \pi ^4 (N+124)+16 (107789 N-904494)+42 \pi ^2 (799 N+9576)\right) \epsilon ^3}{16128 (3 N+22)^3}\\
        &+\sum^{\infty}_{\ell=1}\frac{q_{\perp,\ell}}{36 (\ell^2 (1 + 2 \ell)^5 (-3 + \ell + 4 \ell^2 (2 + \ell))^2 (22 + 3 N)^3)}\Big(8807680 \ell ^{13}\\
        &+55094144 \ell ^{12}+136795904 \ell ^{11}+163417408 \ell ^{10}+75804960
   \ell ^9-30190776 \ell ^8\\
   &-51242800 \ell ^7-18220160 \ell ^6+2647645 \ell ^5+3492167
   \ell ^4+828471 \ell ^3+46917 \ell ^2-2160 \ell +162\Big)\\
                    &+\sum^{\infty}_{\ell=1}\frac{q_{||,\ell}}{36 \ell^2 (1 + 2 \ell)^5 (-3 + \ell + 4 \ell^2 (2 + \ell))^2 (22 + 3 N)^3}\Big(-8807680 \ell ^{13}-46471040 \ell ^{12}\\
                    &-85057280 \ell ^{11}-38597440 \ell ^{10}+74024160
   \ell ^9+110264280 \ell ^8+41686960 \ell ^7-19934320 \ell ^6\\
   &-23508205 \ell ^5-6640865
   \ell ^4+1397445 \ell ^3+1338885 \ell ^2+224100 \ell +20250\Big) \ ,
    \end{split}
\end{equation}
\begin{equation}
\begin{split}
    c_{14}&=\frac{1}{3870720 (3 N+22)^4}\Big(-8745702208 N-51030 \pi ^6 (N+624)-1063125 \pi ^4 (3 N+112)\\
    &-7560 \pi ^2 (20401 N+233624)+77069342208\Big) \    \\
    &+\sum_{\ell=2}^{\infty}\frac{q_{\perp,\ell}}{192 \ell^3 (1 + \ell)^3 (-3 + 2 \ell) (1 + 2 \ell)^7 (5 + 2 \ell) (-2 + \ell + 
   \ell^2) (-3 + 4 \ell (1 + \ell))^3 (22 + 3 N)^4}\Big(97200\\
   &- 1338120 \ell + 15035760 \ell^2 - 
 411372216 \ell^3 - 9774603306 \ell^4 - 
 57869828109 \ell^5 - 91858739832 \ell^6\\
 &+ 
 357767104900 \ell^7 + 1702195436428 \ell^8 + 
 1584847713693 \ell^9 - 4989962083786 \ell^{10} - 
 14377864281052 \ell^{11}\\
 &- 7753331848488 \ell^{12} + 
 21698833645456 \ell^{13} + 41591441475808 \ell^{14} + 
 19449241904448 \ell^{15} \\
 &- 21650115867904 \ell^{16} - 
 34614300248320 \ell^{17} - 16055910612480 \ell^{18} + 
 2733748587520 \ell^{19}\\
 &+ 6553318881280 \ell^{20} + 
 3163530670080 \ell^{21} + 704044113920 \ell^{22} + 
 62538629120 \ell^{23}\Big)\\
                    &+\sum^{\infty}_{\ell=2}\frac{q_{||,\ell}}{192 \ell^3 (1 + \ell)^3 (-3 + 2 \ell) (1 + 2 \ell)^7 (5 + 2 \ell) (-2 + \ell + 
   \ell^2) (-3 + 4 \ell (1 + \ell))^3 (22 + 3 N)^4}\Big(60750000\\
   &+ 796635000 \ell + 4416822000 \ell^2 + 
 16680659400 \ell^3 + 8212333590 \ell^4 - 
 142540734285 \ell^5 \\
 &- 544480094520 \ell^6 - 
 596911459580 \ell^7 + 1776166491340 \ell^8 + 
 6628560684765 \ell^9 + 5086132986230 \ell^{10} \\
 &- 
 11956512418780 \ell^{11} - 28203153835560 \ell^{12} - 
 11996867726960 \ell^{13} + 27520868435680 \ell^{14}\\
 &+ 
 39098158720320 \ell^{15}+ 8444393450240 \ell^{16}- 
 20450538031360 \ell^{17} - 18365446110720 \ell^{18}\\
 &- 
 3158789780480 \ell^{19}+ 3630946416640 \ell^{20} + 
 2496925347840 \ell^{21} + 643443630080 \ell^{22} + 
 62538629120 \ell^{23}\Big)\\
 &+q_{\perp,\ell=1}\frac{78201023 }{9450 (22 + 3 N)^4} \ .
\end{split}
\end{equation}

\end{document}